\documentclass[preprint,english,showpacs,12pt,floatfix]{revtex4}
\usepackage{babel,amsmath,amssymb,dcolumn}
\usepackage{hyperref}
\usepackage{amsfonts}
\usepackage{amssymb}
\usepackage{graphicx}
\usepackage{latexsym}
\usepackage{dcolumn}
\usepackage{epsfig}
\usepackage{subfigure}
\begin{document}

\newcommand{\vp}{\varphi}
\newcommand{\nn}{\nonumber\\}
\newcommand{\beq}{\begin{equation}}
\newcommand{\eeq}{\end{equation}}
\newcommand{\bed}{\begin{displaymath}}
\newcommand{\eed}{\end{displaymath}}
\def\bea{\begin{eqnarray}}
\def\eea{\end{eqnarray}}
\newcommand{\veps}{\varepsilon}
\newcommand{\nablasl}{{\slash \negthinspace \negthinspace \negthinspace \negthinspace  \nabla}}
\newcommand{\om}{\omega}

\newcommand{\Dsl}{{\slash \negthinspace \negthinspace \negthinspace \negthinspace  D}}
\newcommand{\tDsl}{{\tilde \Dsl}}
\newcommand{\tnablasl}{{\tilde \nablasl}}
\title{Scalar field propagation in higher dimensional black holes at a Lifshitz point}

\author{Elcio Abdalla$^{1}$ , Owen Pavel Fern\'{a}ndez Piedra$^{1, 2}$ , Fidel Sosa Nu\~{n}ez$^{2}$, Jeferson de Oliveira$^{3}$ \\
\textit{$^1$Instituto de F\'{\i}sica, Universidade de S\~ao Paulo,
  CP 66318,
05315-970, S\~ao Paulo, Brazil.}\\
\textit{$^2$Grupo de Estudios Avanzados, Universidad de Cienfuegos, Carretera a Rodas, Cuatro Caminos, s/n. Cienfuegos, Cuba.}\\
\textit{$^3$ Instituto de F\'isica, Universidade Federal do Mato Grosso,
CEP 78060-900, Cuiab\'a, Brazil }}
\email{eabdalla@if.usp.br, opavel@ucf.edu.cu, fsosa@ucf.edu.cu, jeferson@fisica.ufmt.br}

\begin{abstract}
We study the complete time evolution of scalar fields propagating in space-times of higher dimensional Lifshitz
Black Holes with dynamical critical exponent $z=2$, obtained from a theory including the most general quadratic
curvature corrections to Einstein-Hilbert gravity in $D$ dimensions. We also computed the quasinormal spectrum after performing a numerical integration and solving exactly the Klein-Gordon equation obeyed by the massive scalar field. We found that quasinormal modes are purely imaginary for all dimensions.

\end{abstract}
\preprint{GEA-UCF 2012-03}
\date{\today}
\maketitle

\section{Introduction}
The study of black hole perturbations has been of interest since the original work of Regge and Wheeler \cite{Regge:1957td}.
The consideration of the fluctuations of the background space-time geometry is related to experiments for the detection of
gravitational waves from supernovae or from coalescence of binary neutron stars, that are supposed to be candidates for the formation
of black holes. During this process, gravitational waves are emited, a process that at late times is dominated by the
fundamental quasinormal mode \cite{Kokkotasreview}.

Apart from gravitational fluctuations, it is useful to consider perturbations of matter fields with different spin weights outside
the Black Hole space-times. It is know that the time evolution of such perturbations can in general be divided in three stages:
a short period of initial outburst of the perturbation, followed by a long period characterized by proper damping quasinormal
oscillations, and at late times, depending of the asymptotic of the space-time at spatial infinity, a stage in which quasinormal
modes are suppressed by power-law or exponential tails \cite{gpp,Abdalla:2008te,Nollertreview,bertireview,zhidenkoreview}.

Quasinormal modes are entirely fixed by the geometry of the background Black Hole space-time and are independent of the initial
conditions on the perturbations. There are similarities with the usual normal modes of closed systems, but the open nature of
the physical region related to the fact that the fluctuations can be absorbed by black hole or radiated to infinity, lead to
frequencies that in general are complex, and do not form a complete set, dominating the evolution over a long but well defined
period of time. As quasinormal frequencies depends on the parameters of the black hole geometry, their computations can serve
to estimate such parameters \cite{berti3}.

On the other hand, the well known relationship between Anti de Sitter (AdS) bulk space gravitational problem in $D$ dimensions and
Conformal Field
Theory (CFT) at the $(D-1)$-dimensional border, also known as AdS/CFT relation, extrapolated its original applications to be studied
in many areas, including phase transitions in the CFT side and generalizations of gravity in the bulk. Quasinormal modes are
also important in this context \cite{maldacenaadscft,aharony,Gubser:2009md}. Related to this idea, a black hole in
asymptotically AdS space-time is dual to a strongly interacting system defined at the AdS boundary, and the imaginary
part of the fundamental quasinormal mode is proportional to the thermalization time in the boundary theory
\cite{horowitz-hubeny,kovtun-starinets}. The gauge/gravity duality has been applied to study a great variety of physical
systems, from quark-gluon plasmas to holographic superconductors \cite{Son:2007vk,Hartnoll:2009sz}.

Recently, gravity solutions with strong asymmetry of time with respect to space coordinates have been obtained, both in
the severely modified Horava Lifshitz gravity \cite{myung} as well as in Einstein gravity with
higher order curvature corrections \cite{eloyhdim}.

Several theories present anisotropic scaling when one compares time with space coordinates. Such rescaling appears
frequently when higher powers of curvature are present in the gravity action. The so called
$D$-dimensional Lifshitz backgrounds have the form
\begin{equation}
ds^{2}=-\frac{r^{2z}}{L^{2z}}dt^{2}+\frac{l^{2}}{r^{2}}dr^{2}
+\frac{r^{2}}{L^{2}}\sum_{i=1}^{D-2}dx^adx^a\quad ,  \label{lif}
\end{equation}
where $x^a$ are the components of a $(D-2)$-dimensional coordinate vector. The geometry described by the above line element admits
the anisotropic scaling symmetry
\begin{equation}\label{scaling}
t\mapsto\lambda^{z}t\quad ,\qquad r\mapsto\lambda^{-1}r\quad ,\qquad
x^a\mapsto\lambda x^a\quad ,
\end{equation}
which is precisely part of the isometry group present in their non-relativistic duals, that are considered on the $(D-1)$-dimensional boundary at infinity.

This paper is the first of a series in which a complete study of the time evolution of perturbations around Black Hole
solutions whose asymptotic is the space-time (\ref{lif}) is performed. We will consider scalar perturbations, that are important not only due to its simplicity, but for
the implementation of gauge/gravity duality ideas. Our motivation is basically concerned with the extension of the AdS/CFT
correspondence to other areas of physics, including further condensed matter models. Within the context of non-relativistic
physics, gauge/gravity ideas were recently addressed, as in references \cite{Son} and \cite{balasubramanian}, where the authors proposed gravity duals to non-relativistic systems, and in \cite{Kachru}, where new gravitational duals to Lifshitz fixed points with scale invariance but without Galilean invariance were presented.

Previously, the three dimensional case had been considered, the so called New Massive Gravity\cite{bertha}, and it has been
conjectured that such a theory is dual to the KdV equation in two dimensional space-time\cite{abdallaoliveira}. It will be
encouraging if one is able to find similar relations in higher dimensions.

The paper is organized as follows: Section II briefly presents the line element of higher dimensional asymptotically
Lifshitz black holes obtained considering the most general quadratic corrections to Einstein gravity. In Section III
we obtain a general wave equation suitable to analyze the propagation of scalar fields in the
background geometry of the higher dimensional black holes, whereas Section IV is devoted to the numerical solution of the
evolution equations, and the computation of the complex quasinormal frequencies using two methods: fitting time
domain data and the Horowitz-Hubbeny approach. We also present, in Section V, the exact results for quasinormal
frequencies after solving the equation of motion for the scalar field in the general $D$-dimensional Lifshitz
backgrounds. Finally, we present our conclusions.

\section{$D\geq5$ asymptotically Lifshitz black hole families from quadratic corrected gravity}

In this section we briefly present three different classes
of asymptotically Lifshitz black hole solutions which
correspond to different ranges of the dynamical exponent $z$ \cite{eloyhdim}.
As was shown in \cite{eloyhdim} below, there exist at least one family of black hole solutions for any generic value of the dynamical exponent $z$.

In the following we will consider the action:
\begin{eqnarray}
S&=&\int{d}^Dx\sqrt{-g} \left(R-2\Lambda+\xi_1{R}^2
+\xi_2{R}_{\mu\nu}{R}^{\mu\nu}
+\xi_3{R}_{\mu\nu\sigma\tau}{R}^{\mu\nu\sigma\tau} \right)\quad .
\label{e1}
\end{eqnarray}
that includes the most
general quadratic-curvature corrections to $D$-dimensional gravity, and we will focus only in the higher dimensional cases with $D\geq5$.

The field equations that arise from the action (\ref{e1}) are \cite{eloyhdim}:
\begin{eqnarray}
G_{\mu\nu}+\Lambda{g}_{\mu\nu}
+\vartheta_{1}\square{R}_{\mu\nu}
+\vartheta_{2}g_{\mu\nu}\square{R}
+\vartheta_{3}\nabla_\mu\nabla_\nu{R}+\vartheta_{4}R_{\mu\sigma\nu\tau}R^{\sigma\tau}+2\xi_1RR_{\mu\nu}
\nonumber\\ \nonumber\\
{}+2\xi_3\left(R_{\mu\gamma\sigma\tau}R_{\nu}^{~\gamma\sigma\tau}
-2R_{\mu\sigma}R_{\nu}^{~\sigma}\right)-\frac12\left(\xi_1{R}^2+\xi_2{R}_{\sigma\tau}{R}^{\sigma\tau}
+\xi_3{R}_{\sigma\tau\gamma\delta}{R}^{\sigma\tau\gamma\delta}
\right)g_{\mu\nu}&=&0\quad .\qquad \label{e2}
\end{eqnarray}
where the coefficients $\vartheta_{k}, \ \ k=1,...,4$ in front of some geometric factors are functions of those appearing accompanying the quadratic terms in the action (\ref{e1}) and are given by:
\begin{eqnarray}\label{coequ}
\vartheta_{1}&=&\xi_2+4\xi_3  \,\\
\vartheta_{2}&=&\frac{1}{2}\left(4\xi_1+\xi_2\right) \,\\
\vartheta_{3}&=& -\left(2\xi_1+\xi_2+2\xi_3\right) \,\\
\vartheta_{4}&=&2\left(\xi_2+2\xi_3\right)\,\label{coef2}
\end{eqnarray}
The above field equations have solutions whose line elements are particular cases of those describing a \(D\)-dimensional manifold
\(\mathcal{M}\), and that can be written locally as a warped product of the form
\begin{equation}\label{warpedproduct1}
    \mathcal{M}^{D}=\mathcal{O}^{2}\times\mathcal{K}^{D-2}\quad ,
\end{equation}
where we assume that $\mathcal{N}^{m}$ is a lorentzian and \(\mathcal{K}^{D-2}\)  an Einstein manifold.
Introducing coordinates $z^{\mu}=(y^{a},x^{i})$ in $\mathcal{M}^{D}$, where $y^{a}$ and $x^{i}$ are coordinates defined in the manifolds
$\mathcal{O}^{2}$ and $\mathcal{K}^{D-2}$, respectively, the full metric describing the geometry of the manifold $\mathcal{M}^{D}$ can
be written as
\begin{equation}\label{warpedmetric2}
    ds^{2}=\mathfrak{g}_{\mu\nu}dz^{\mu}dz^{\nu}=g_{ab}(y^{a})dy^{a}dy^{b}+w^{2}(y^{a})d\sigma_{D-2}^{2}\quad ,
\end{equation}
where $w(y^{a})$ is an arbitrary function defined in $\mathcal{O}^{2}$ and
\begin{equation}\label{maxspacemetric1}
    d\sigma_{n}^{2}=\gamma_{ij}(x)dx^{i}dx^{j}
\end{equation}
is the metric on \(\mathcal{K}^{D-2}\).

A common realization of (\ref{warpedmetric2}) is that in which we chose radial and time coordinates $(y^{1},y^{2})\equiv(r,t)$
in $\mathcal{O}^{2}$ and describe the geometry of $\mathcal{M}^{D}$ by the line element
\begin{equation}\label{metricgral}
ds^{2}=-A(r)dt^{2}+B(r)dr^{2}+C(r)d\Sigma_{D-2}^{2}(x^{i})\quad .
\end{equation}

As an example we can mention the Lifshitz spacetimes (\ref{lif}), that are solutions of (\ref{e2}) for a general
value of the dynamical exponent $z$ in any dimension, in such a way that they exist for fixed values of the cosmological constant and the parameter $\beta_{2}$ given by
\begin{eqnarray}
\Lambda &=&
-\frac1{4L^2}\bigg(2z^2+(D-2)(2z+D-1)-\frac{4(D-3)(D-4)z(z+D-2)\beta_3}{L^2}\bigg)\quad ,
\label{eq:Lparam_lambda}\\\nonumber\\
\beta_2 &=& \frac{L^2-2\left[2z^2+(D-2)(2z+D-1)\right]\beta_1
-4\left[z^2-(D-2)z+1\right]\beta_3}
{2(z^2+D-2)}\quad ,\label{param1}
\end{eqnarray}
generalizing the results for the $D=3$ New Massive Gravity obtained in Ref. \cite{AyonBeatonmg} for $\beta_3=0$ and
$\beta_2=-(8/3)\beta_1=-1/m^2$. The formula (\ref{lif}) can be written in the form (\ref{metricgral}) with
$A(r)=\frac{r^{2z}}{L^{2z}}$, $B(r)=\frac{L^{2}}{r^{2}}$, $C(r)=\frac{r^{2}}{L^{2}}$ and $\gamma_{ij}(x^{i})=diag[1,1,...,1], \ i,j=1,..d-2$.

If the line element (\ref{warpedmetric2}) describes a black-hole space-time (that is not the case of (\ref{lif})), then
\(\mathcal{K}^{n}\) describes the structure of a spatial section of its event horizon. If \(\mathcal{K}^{n}\) is a
constant curvature space with sectional curvature \(K\), then \(K=0\) represents a flat space, \(K=1\) a spherical
manifold and  \(K=-1\) a hyperbolic one \cite{ik}.

For $D\geq5$ there are three different Lifshitz black hole families. The geometry of all these families are described by
(\ref{metricgral}) with $C(r)=\frac{r^2}{L^2}$ and $d\Sigma_{D-2}^{2}(x^{i})\equiv dx^adx^a, \ \ a=1,..D-2$ represents a flat
euclidean $(D-2)$-dimensional manifold with \(K=0\). The first family of solutions describes asymptotically Lifshitz
black holes for the dynamical exponent $z>2-D$, and its line element is given by (\ref{metricgral}) with
\begin{equation}\label{lbh1}
A(r)=\frac{r^{2z}}{L^{2z}}\left(1-\frac{ML^{(z+D-2)/2}}{r^{(z+D-2)/2}}\right)\quad , \ \
B(r)=\frac{L^2}{r^2}\left(1-\frac{ML^{(z+D-2)/2}}{r^{(z+D-2)/2}}\right)^{-1}\quad .
\end{equation}
The dynamical exponent $z$ parameterizes the coupling constants of the theory. This solution is defined only for $D\ge5$, and there exists a
conformal limit $z=1$ in the form of an asymptotically AdS black hole that coincides with a solution
of the Einstein-Gauss-Bonnet gravity with a fine-tuned coupling constant.

The second family of asymptotically Lifshitz black holes are obtained for $z>1$, and are of the form (\ref{metricgral}) with
\begin{equation}\label{lbh2}
A(r)=\frac{r^{2z}}{L^{2z}}
\left(1-\frac{ML^{2(z-1)}}{r^{2(z-1)}}\right)\quad , \ \
B(r)=\frac{L^2}{r^2}
\left(1-\frac{ML^{2(z-1)}}{r^{2(z-1)}}\right)^{-1}\quad ,
\end{equation}
and is compatible with the coupling constants given by
\begin{eqnarray*}
\Lambda&=&-\frac{(z-1)\left[z^2-Dz-(D-1)(D-2)\right]}{2L^2(z-D)}\quad ,\\
\nonumber\\
\beta_1&=&L^2\Big[3(D-1)(D-2)z^3
-(2D^3-2D^2-11D+20)z^2+(3D^3-14D^2+19D+10)z\\
\nonumber\\
&&{}+(D+2)(D-4)\Big]\Big/{\Big[}2(D-2)(D-3)(D-4)z(z-1)(z-D)(3z+D-4)
{\Big]}\quad ,\\
\nonumber\\
\beta_2&=&-L^2(D-1)(2z-D-2)\frac{[6(D-2)z^2-(D^2-3D+8)z-2(D-4)]}{[2(D-2)(D-3)(D-4)z(z-1)(z-D)(3z+D-4)]}\quad ,\\
\nonumber\\
\beta_3&=&\frac{L^2(D-1)(2z-D-2)}{4(D-3)(D-4)z(z-D)}\quad .
\end{eqnarray*}
In similarity with the above case for the $z>2-D$ black hole family, the solution exists only for higher dimensions $D\geq5$ and a conformal limit $z=1$ is absent.

In reference \cite{eloyhdim} the authors also found a third family of Lifshitz black holes caracterized by a negative
dynamical critical exponent $z=-|z|$, with metric given by (\ref{metricgral}) with
\begin{equation}\label{lbh3}
A(r)=\frac{L^{2|z|}}{r^{2|z|}}
\left(1-\frac{ML^{|z|}}{r^{|z|}}\right)\quad , \ \
B(r)=\frac{L^2}{r^2} \left(1-\frac{ML^{|z|}}{r^{|z|}}\right)^{-1}\quad ,
\end{equation}
with the corresponding coupling constants parameterized as usual by $z$. Again this family of black holes is defined only in higher
dimensions $D\ge5$, and due to the presence of a negative dynamical
critical exponents, it has no conformal limit $z=1$.

In the following we will be interest in the second family of higher dimensional Lifshitz black holes discussed above and
specifically those solutions that are caracterized by the dynamical exponent $z=2$.


\section{Causal Structure of the $z=2$ Lifshitz Black Hole }

In this section we study the causal structure of $z=2$ asymptotically Lifshitz black holes. Considering for simplicity the five dimensional case and taking $L=1$, we can write the line element as
\begin{equation}\label{metric_z2}
ds^{2}=-r^{4}\left(1-\frac{r_{+}^{2}}{r^{2}}\right)dt^{2}+\frac{1}{r^{2}\left(1-\frac{r_{+}^{2}}{r^{2}}
\right)}dr^{2}+r^{2}\left(dx^{2}_{1}+dx^{2}_2+dx^{2}_{3}\right)\quad .
\end{equation}

Regarding to the causal structure, the above metric is very similar to the $z=3$ NMG black hole \cite{bertha}.
There is a single regular event horizon at $r_{+}=\sqrt{M}$, where $M$ is an integration constant related to de ADM
mass. Also, the metric exhibits a spacetime singularity at $r=0$, as we can see from the behavior of the Kretschmann invariant
\begin{equation}\label{kinv}
R_{\mu\nu\sigma\rho}R^{\mu\nu\sigma\rho}=-\frac{8}{r^4}\left(3r_{+}^{4}-9r^{2}_{+}r^{2}+17r^{4}\right)\quad .
\end{equation}
As $r\rightarrow 0$ we have
\[
R_{\mu\nu\sigma\rho}R^{\mu\nu\sigma\rho}\rightarrow \infty\quad .
\]
We also notice that as $r\rightarrow r_{+}$
\[
R_{\mu\nu\sigma\rho}R^{\mu\nu\sigma\rho}\rightarrow -88\quad .
\]
Therefore, from these facts we can say that there is a genuine physical singularity at $r=0$ and a regular surface at $r=r_{+}$.
Further details can be extracted from the Penrose-Carter Diagram. In order to construct such a diagram, we have to remove the
singularity in the coordinate patch at $r=r_{+}$ through the null coordinates
\[
U=e^{r_{+}^{2}(t+r_{*})}, \hspace{0.3cm} V=-e^{-r_{+}^{2}(t-r_{*})}\quad ,
\]
where $r_{*}$ is the tortoise coordinate given by
\[
r_{*}(r)=\ln\left(1-\frac{r_{+}^{2}}{r^{2}}\right)^{\frac{1}{2r_{+}^{2}}}\quad .
\]
The metric (\ref{metric_z2}) can be rewritten as
\[
ds^{2}=-\left(\frac{r_{+}}{r}\right)^{4}dUdV+r^{2}\left(dx_{1}^{2}+dx_{2}^{2}+dx_{3}^{2}\right)\quad ,
\]
which is obviously regular at $r=r_{+}$.

\begin{figure}[htbp!]
\begin{center}
\includegraphics[height=9cm, width=8cm]{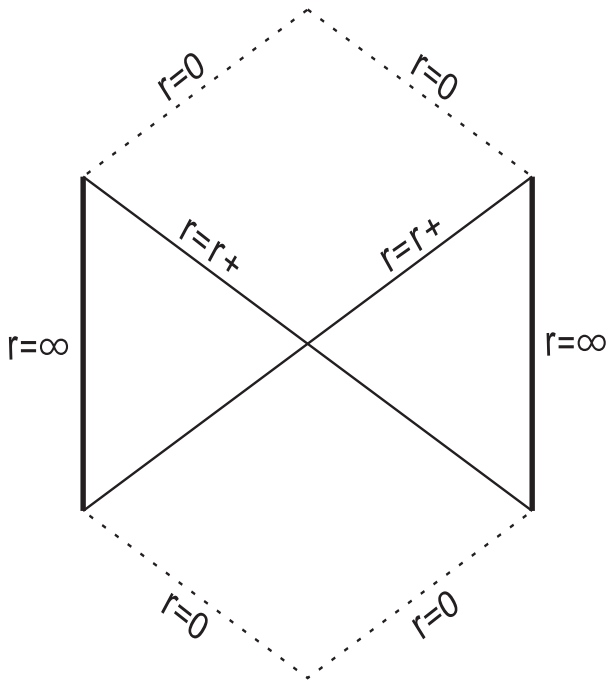}
\caption{Penrose-Carter Diagram for  five dimensional $z=2$ Lifshitz Black Hole. The dashed lines refer to the null physical singularity
at $r=0$ covered by a regular event horizon at $r=r_{+}$.}
\label{penrose}
\end{center}
\end{figure}

Performing another change in the coordinates, that is,
\[
T=\arctan(U)+\arctan(V),\hspace{0.3cm} X=\arctan(U)-\arctan(V)\quad ,
\]
we construct the Penrose-Carter diagram (\ref{penrose}).

The diagrams show us the null nature of the singularity and the AdS-like spatial infinity at $r=\infty$. From this, we can see that the
causal structure of $z=2$ Lifshitz black holes are essentially the same as the ordinary AdS black holes. Thus, one can apply the
gauge/gravity duality ideas in the Lifshitz case, in a pattern similar to the AdS case.

\section{Perturbation equations}

In the space-time (\ref{warpedmetric2}) the Klein-Gordon
equation for a minimally coupled massive scalar field $\phi(z^{\mu})$ with mass $m$ can be written as
\begin{equation}\label{kgcap2-1}
\mathcal{\hat{K}}_{E}\phi(z^{\nu})\equiv
    (\hat{\overline{\square}}-m^{2})\phi(z^{\nu})=0\quad .
\end{equation}
We can write the Laplace-Beltrami operator $\mathcal{\hat{K}}_{E}$ in (\ref{kgcap2-1}) as $\hat{\overline{\square}}\equiv
\overline{\nabla}^{\mu}\overline{\nabla}_{\mu}=\hat{\square}+\frac{n}{w}D_{a}w\cdot D^{a}+w^{-2}\hat{\triangle}$, where
$\hat{\square}\equiv D^{a}D_{a}$ and $\hat{\triangle}\equiv \hat{D}^{i}\hat{D}_{i}$ are the Laplace-Beltrami operators
in $\mathcal{N}^{m}$ and $\mathcal{K}^{n}$. In the above formulas $D^{a}$ and $\hat{D}^{i}$ are the usual covariant
derivatives in $\mathcal{N}^{m}$ and $\mathcal{K}^{n}$.

Introducing scalar harmonics $\mathbb{S}_{k_{S}}(x^{i})$ in $\mathcal{K}^{n}$, that are eigenfunctions of $\hat{\triangle}$ defined by
\begin{equation} \label{c9}
(\hat{\triangle}+k^{2}_{E})\mathbb{S}_{\mathbf{\textbf{k}}_{E}}=0,  \ \ \ \ \int_{\mathcal{K}^{n}}\mathbb{S}_{\mathbf{\textbf{k}}_{E}}
\mathbb{S}_{\mathbf{\textbf{k}'}_{E}}=\delta_{\mathbf{\textbf{k}}_{E},\mathbf{\textbf{k}'}_{E}}\quad ,
\end{equation}
we can separate variables to obtain
\begin{equation}\label{varseparation1}
    \phi(y^{a},x^{i})=w^{-\frac{n}{2}}\sum_{\mathbf{k}_{E}}\varphi_{\mathbf{k}_{E}}(y^{a})\mathbb{S}_{\mathbf{k}_{E}}(x^{i})\quad .
    \end{equation}
Putting (\ref{varseparation1}) in (\ref{kgcap2-1}), we obtain the equation for $\phi(y^{a},x^{i})$ on $\mathcal{N}^{m}$,
\begin{equation}\label{kgcap2-2}
   \hat{\square}\varphi_{\mathbf{k}_{E}}(y^{a})+W_{\mathbf{k}_{E}}^{(E)}\varphi_{\mathbf{k}_{E}}(y^{a})=m^{2}\varphi_{\mathbf{k}_{E}}(y^{a})\quad ,
\end{equation}
where $W_{\mathbf{k}_{E}}^{(E)}\equiv W_{\mathbf{k}_{E}}^{(E)}(y^{a})$ is given by the expression
\begin{equation}\label{W1}
    W_{\mathbf{k}_{E}}^{(E)}(y^{a})=\frac{1}{w^{2}}\left[\frac{n(2-n)}{4}(Dw)^{2}-\frac{n}{2}w\hat{\square} w-k_{E}^{2}\right]\quad .
\end{equation}
Above, $k_{E}^{2}$ are the eigenvalues of the scalar harmonics.
Equation (\ref{kgcap2-2}) describes the propagation of scalar perturbations over a generic $(m+n)$-dimensional space-time $\mathcal{M}$.
Considering now the particular case of a $D$-dimensional space-time given by (\ref{metricgral}), with $d\Sigma_{D-2}^{2}(x^{i})$ representing a flat euclidean $(D-2)$-dimensional manifold, then (\ref{kgcap2-2}) results in
\begin{equation}\label{kgspherical1}
     \hat{\square}_{(2)}\varphi_{\mathbf{k}}^{(E)}+W_{\mathbf{k}}^{(E)}(r)\varphi_{\mathbf{k}}^{(E)}=m^{2}\varphi_{\mathbf{\mathbf{k}}}^{(E)}\quad ,
\end{equation}
with $\varphi_{\mathbf{k}}^{(E)}=\varphi_{\mathbf{k}}^{(E)}(t,r)$ and
\begin{eqnarray}\label{Weffectivescalar}
   W_{\mathbf{k}}^{(E)}(r)&=& \frac{1}{C(r)}\left\{\frac{(D-2)(4-D)}{4}F(r)-\frac{D-2}{2}G(r)-k^{2}\right\}\quad ,\\
\label{fg1}
   F(r)&=&g^{rr}\left[\frac{d}{dr}(\sqrt{C(r)})\right]^{2},\ \ G(r)=\sqrt{C(r)}\hat{\square}_{(2)} (\sqrt{C(r)})\quad .
\end{eqnarray}
where $\hat{\square}_{(2)}$ denotes the Laplace-Beltrami in the two-dimensional orbit space with coordinates $(y^{1},y^{2})=(t,r)$.
Using the explicit metric components in the orbit space in (\ref{kgspherical1})-(\ref{fg1}) we obtain
\begin{equation}\label{kgspherical3}
    -\frac{\partial^{2}\varphi_{\mathbf{k}}^{(E)}}{\partial t^{2}}+\left(\frac{A(r)}{B(r)}\right)\frac{\partial^{2}\varphi_{\mathbf{k}}^{(E)}}
{\partial r^{2}}+\frac{1}{2}\left(\frac{A}{B}\right)\left[\frac{d}{dr}\ln \left(\frac{A}{B}\right)\right]
\frac{\partial\varphi_{\mathbf{k}}^{(E)}}{\partial r}+A\left[W_{\mathbf{k}}^{(E)}(r)-m^{2}\right]\varphi_{\mathbf{k}}^{(E)}=0\quad ,
\end{equation}
The functions $F(r)$ and $G(r)$ in $W_{\ell}^{(E)}(r)$ are given by
\begin{equation}\label{fg2}
   W_{\mathbf{k}}^{(E)}(r)=\frac{1}{C(r)}\left\{\frac{(D-2)(6-D)}{16BC}\left(\frac{dC}{dr}\right)^{2}-
   \frac{D-2}{8B}\left[\frac{d}{dr}\ln \left(\frac{A}{B}\right)\right]\frac{dC}{dr}+\frac{1}{2B}\frac{d^{2}C}{dr^{2}}-k^{2}\right\}\quad .
\end{equation}

We introduce in (\ref{kgspherical3}) the tortoise coordinate $r_{*}$ defined by $\frac{d}{dr_{*}}=\sqrt{\frac{A}{B}}\frac{d}{dr}$.
It maps the physical region $r\in\left(r_{+},\infty\right)$ into $r_{*}\in\left(-\infty,0\right)$. Dropping the indices
$\mathbf{k}$ and $(E)$ we finally obtain for the evolution equation
\begin{equation}\label{timeevolescalar}
    -\frac{\partial^{2}\varphi(t,r)}{\partial t^{2}}+\frac{\partial{\varphi(t,r)}}{\partial
    r_{*}^{2}}=V(r)\varphi(t,r)\quad ,
\end{equation}
where
\begin{eqnarray}
 \nonumber
  V(r) &=& \frac{1}{16}\left(\frac{B}{A}\right)\left[4H_{1}^{2}-8H_{1}\frac{d}{dr}\left(\frac{A}{B}\right)+
3\left[\frac{d}{dr}\left(\frac{A}{B}\right)\right]^{2}
    -4\frac{A}{B}\frac{d^{2}}{dr^{2}}\left(\frac{A}{B}\right)+ 8\frac{A}{B}\frac{d}{dr}H_{1}\right]  \\
     &-& A\left[W_{\mathbf{k}}^{(E)}(r)-m^{2}\right]
\end{eqnarray}
plays the role of effective potential.
\section{Time evolution of perturbations}
\begin{figure}[t]
\begin{center}
\scalebox{0.27}{\includegraphics{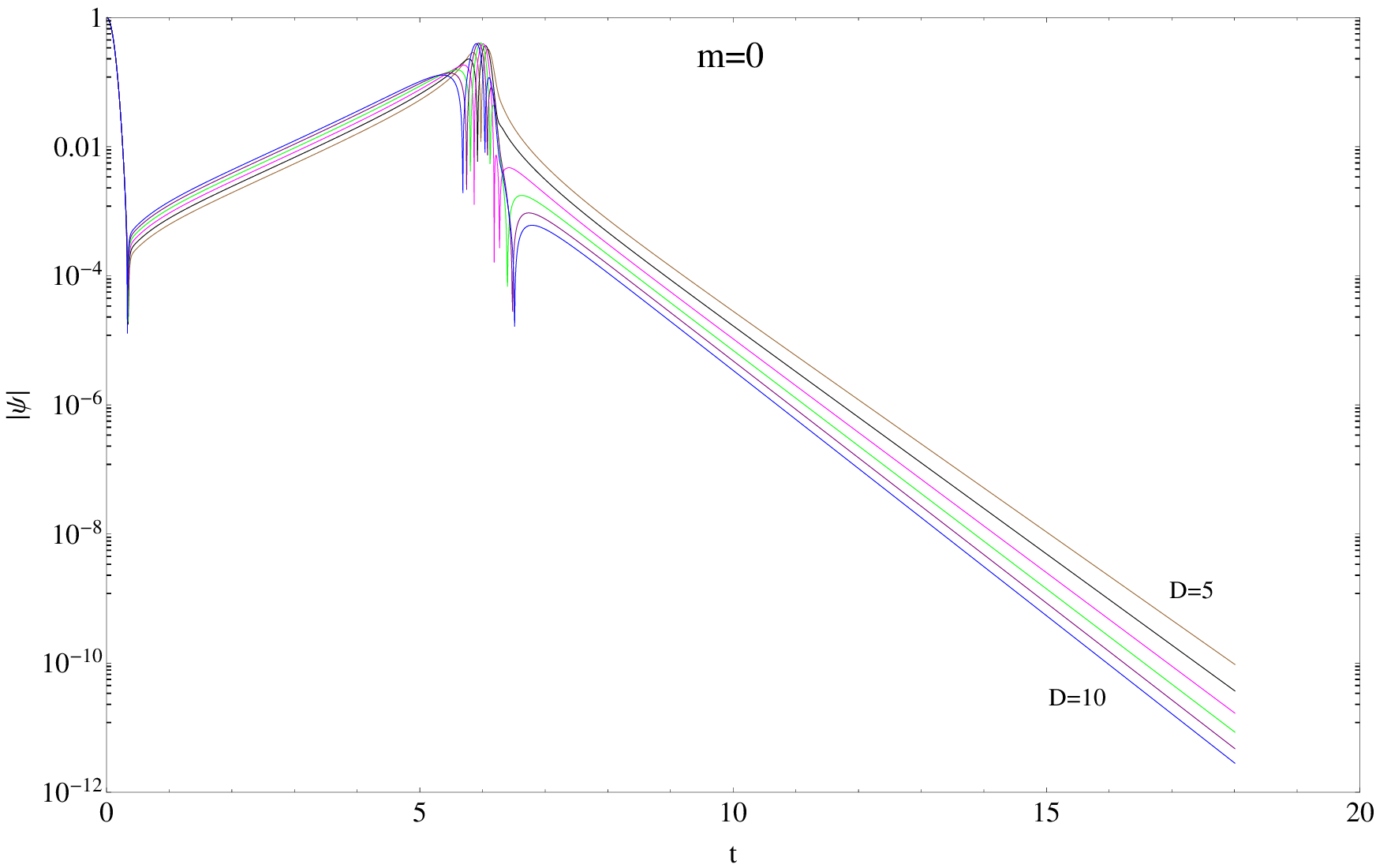}}
\vspace{0.22cm}
\scalebox{0.27}{\includegraphics{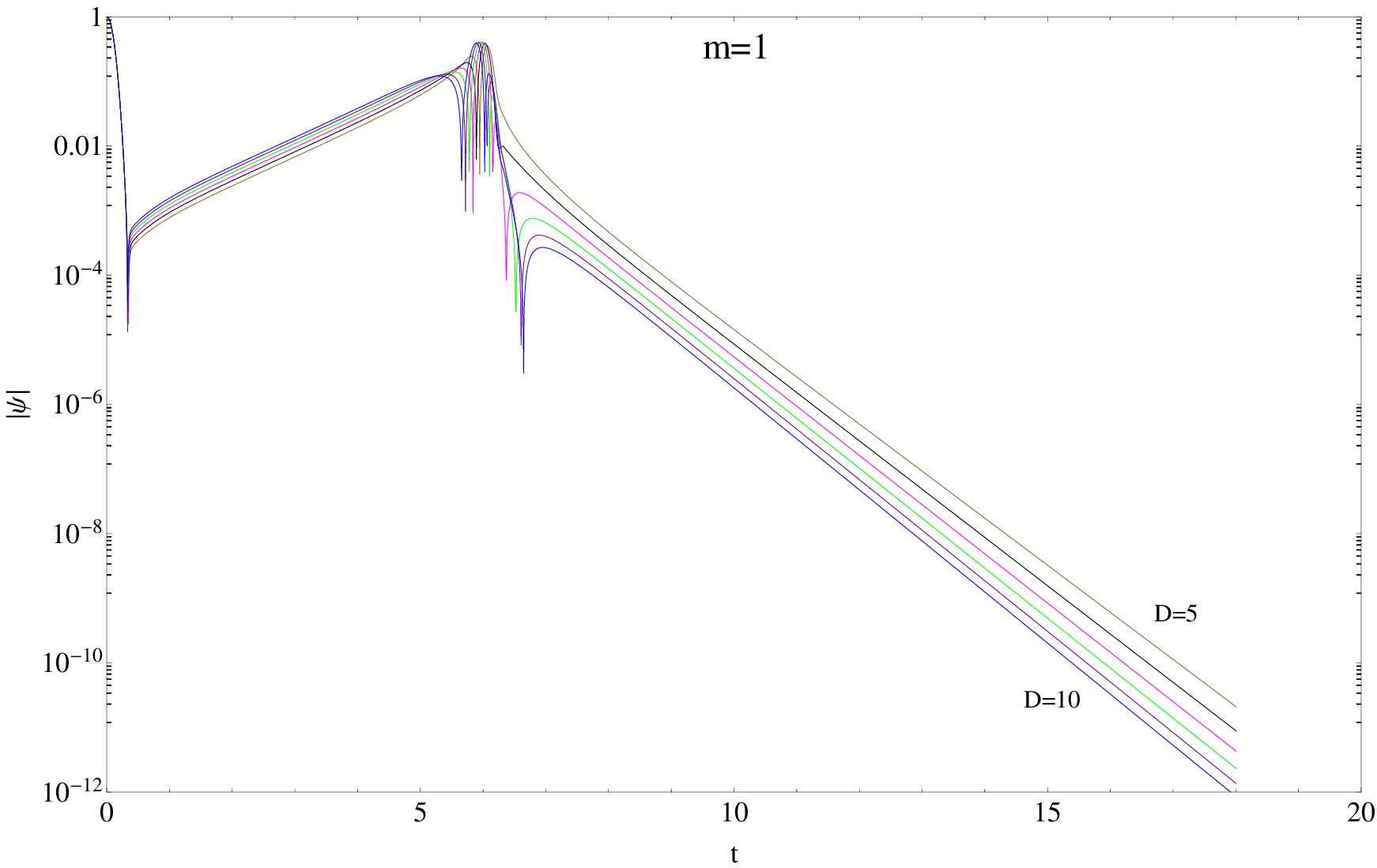}}
\end{center}
\caption{\it Semi-logarithmic plots of the
time-domain evolution of $m=0$ and $m=1$ scalar perturbations in Lifshitz black holes for various dimensions.
The parameters are $M=L=k=1$ and from top to bottom curves $D=5,...,10$.} \label{perfiles1}
\end{figure}
\begin{figure}[t]
\begin{center}
\scalebox{0.27}{\includegraphics{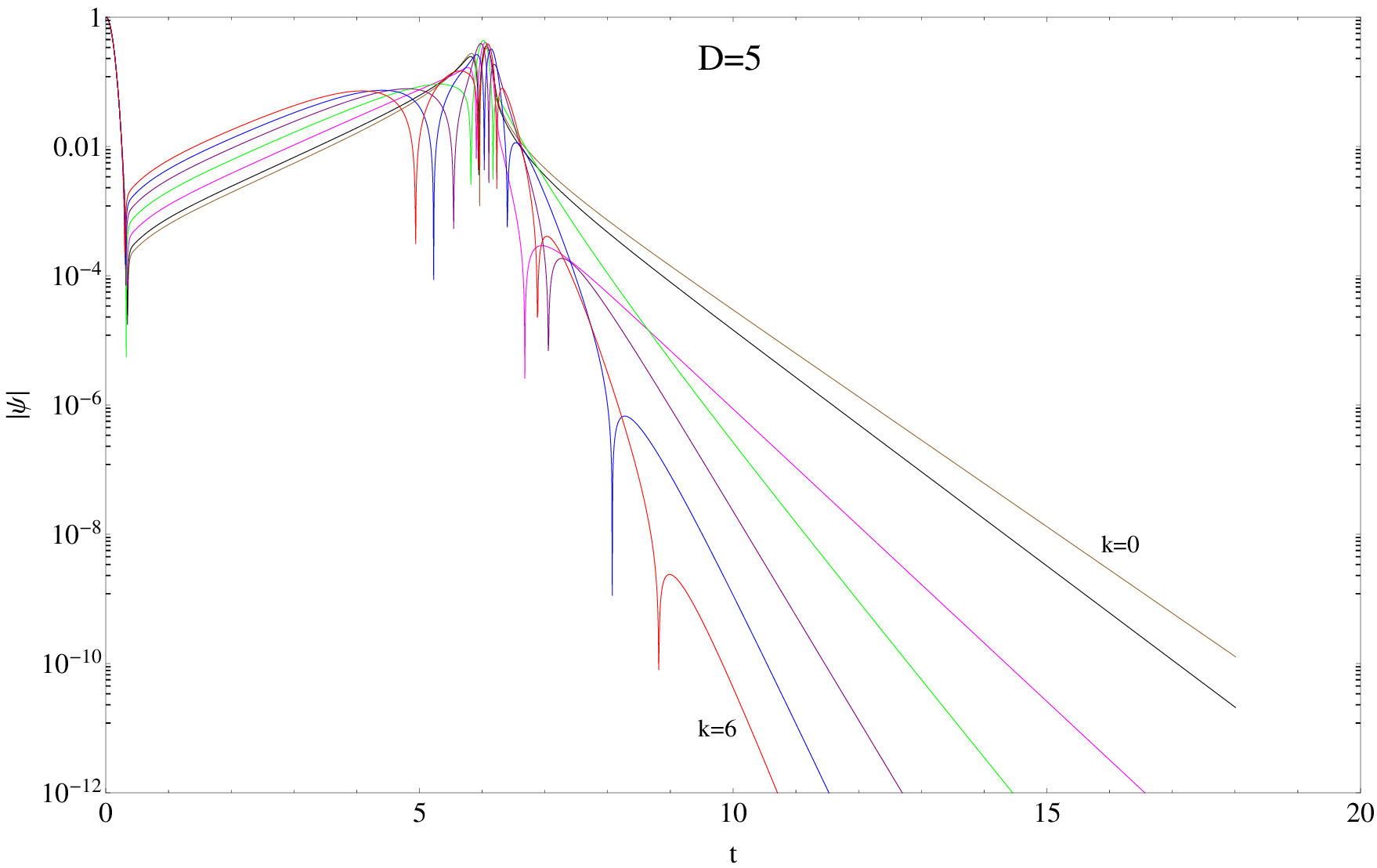}}
\vspace{0.22cm}
\scalebox{0.27}{\includegraphics{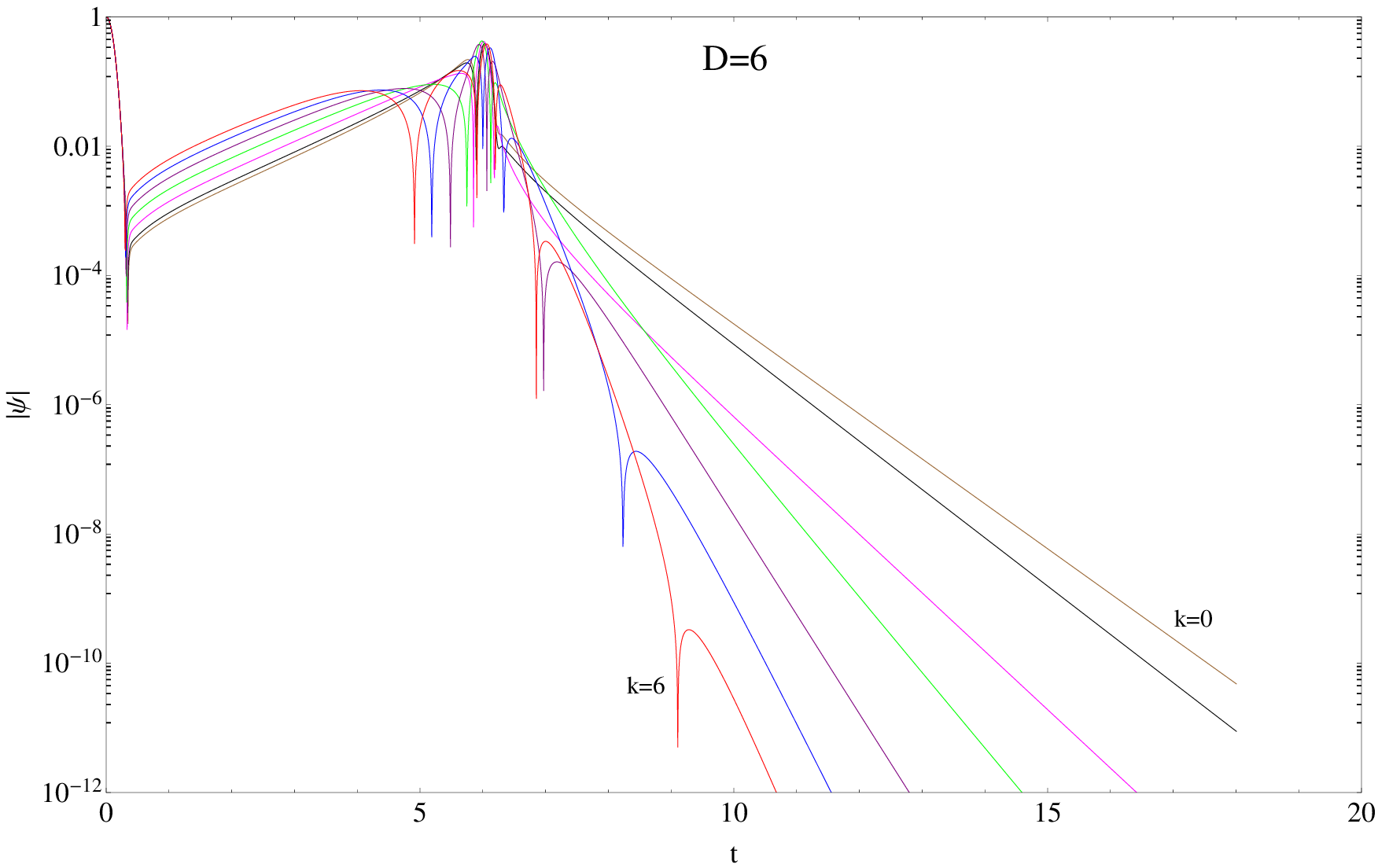}}
\vspace{0.22cm}
\scalebox{0.27}{\includegraphics{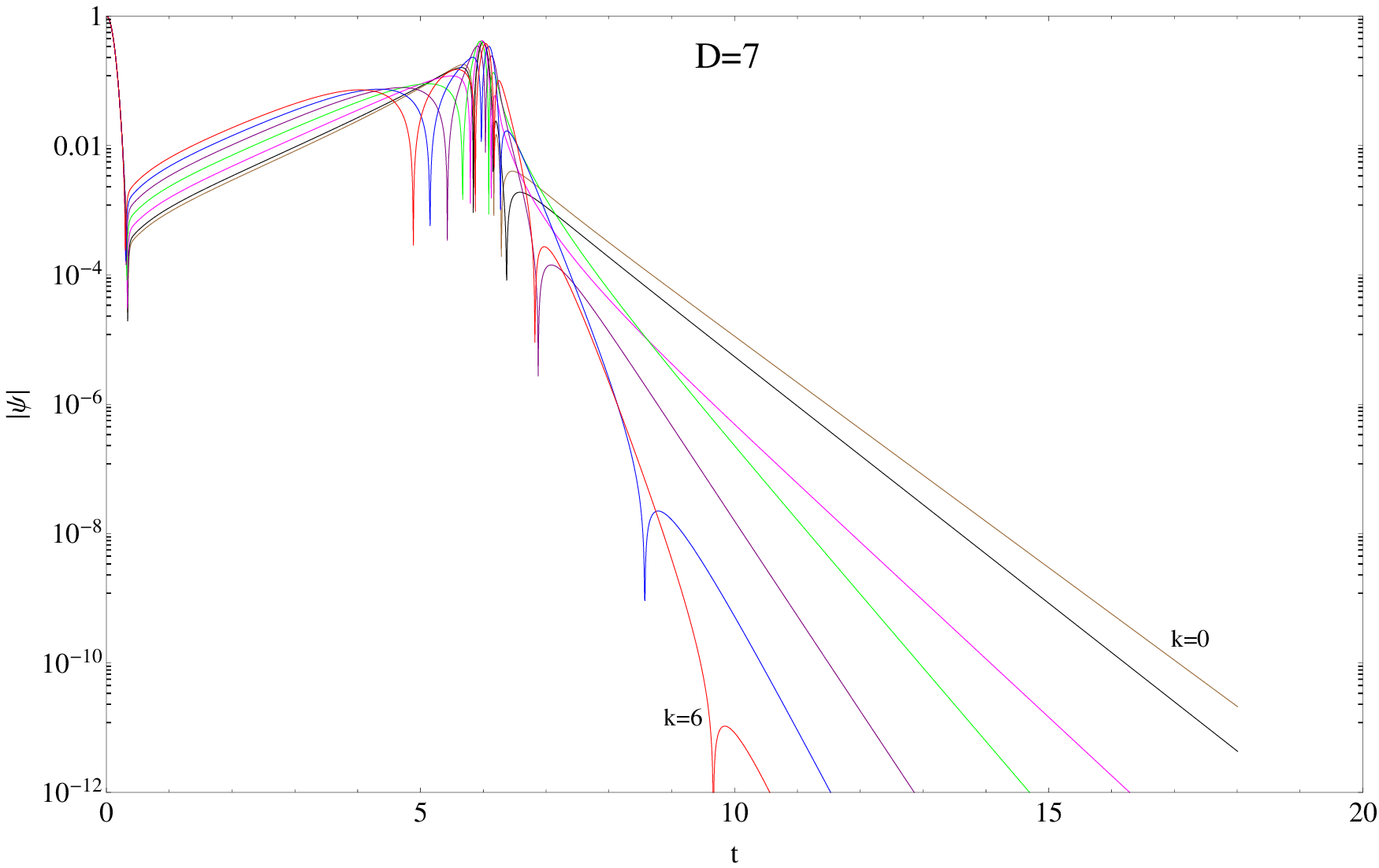}}
\vspace{0.22cm}
\scalebox{0.27}{\includegraphics{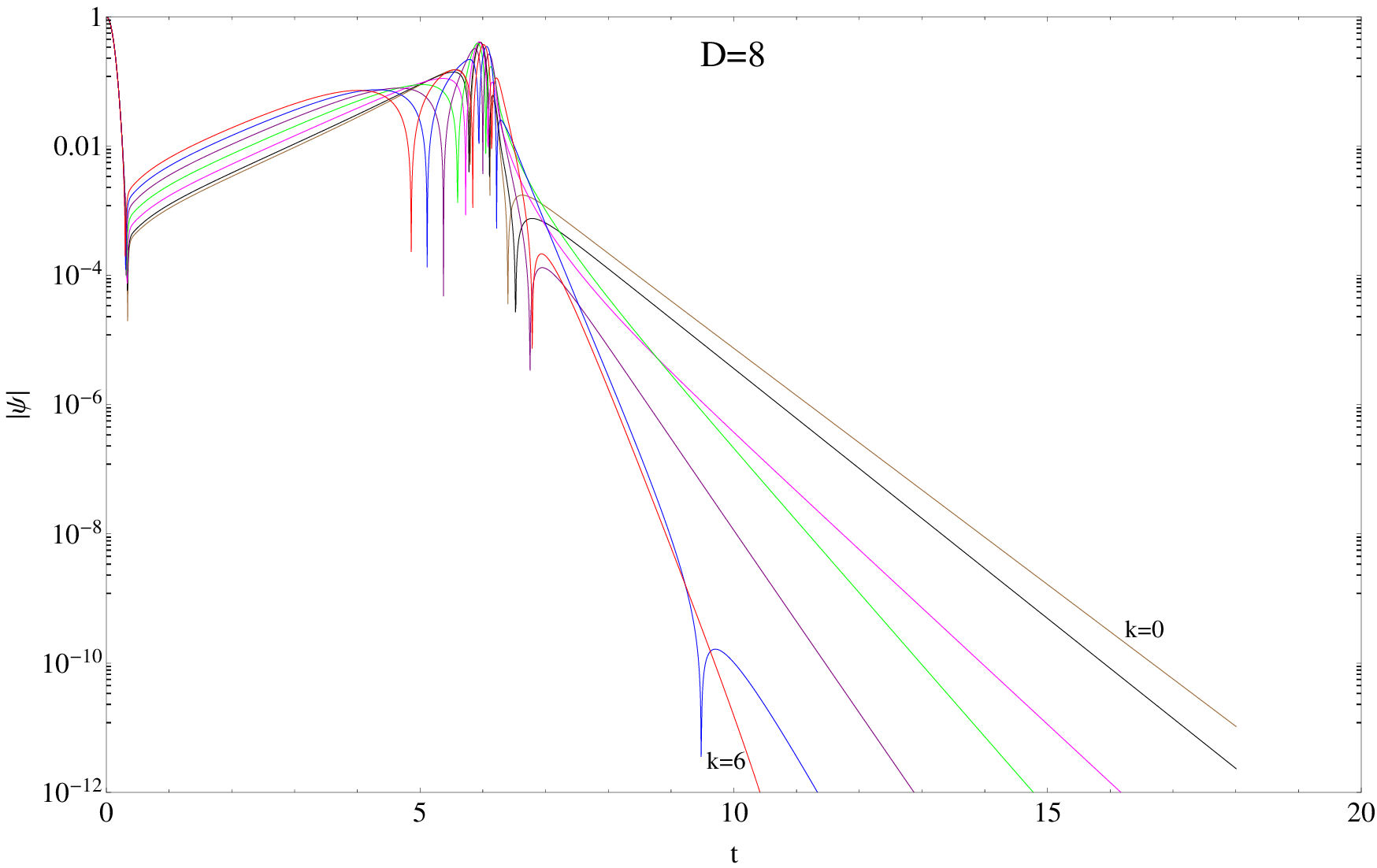}}
\vspace{0.22cm}
\scalebox{0.27}{\includegraphics{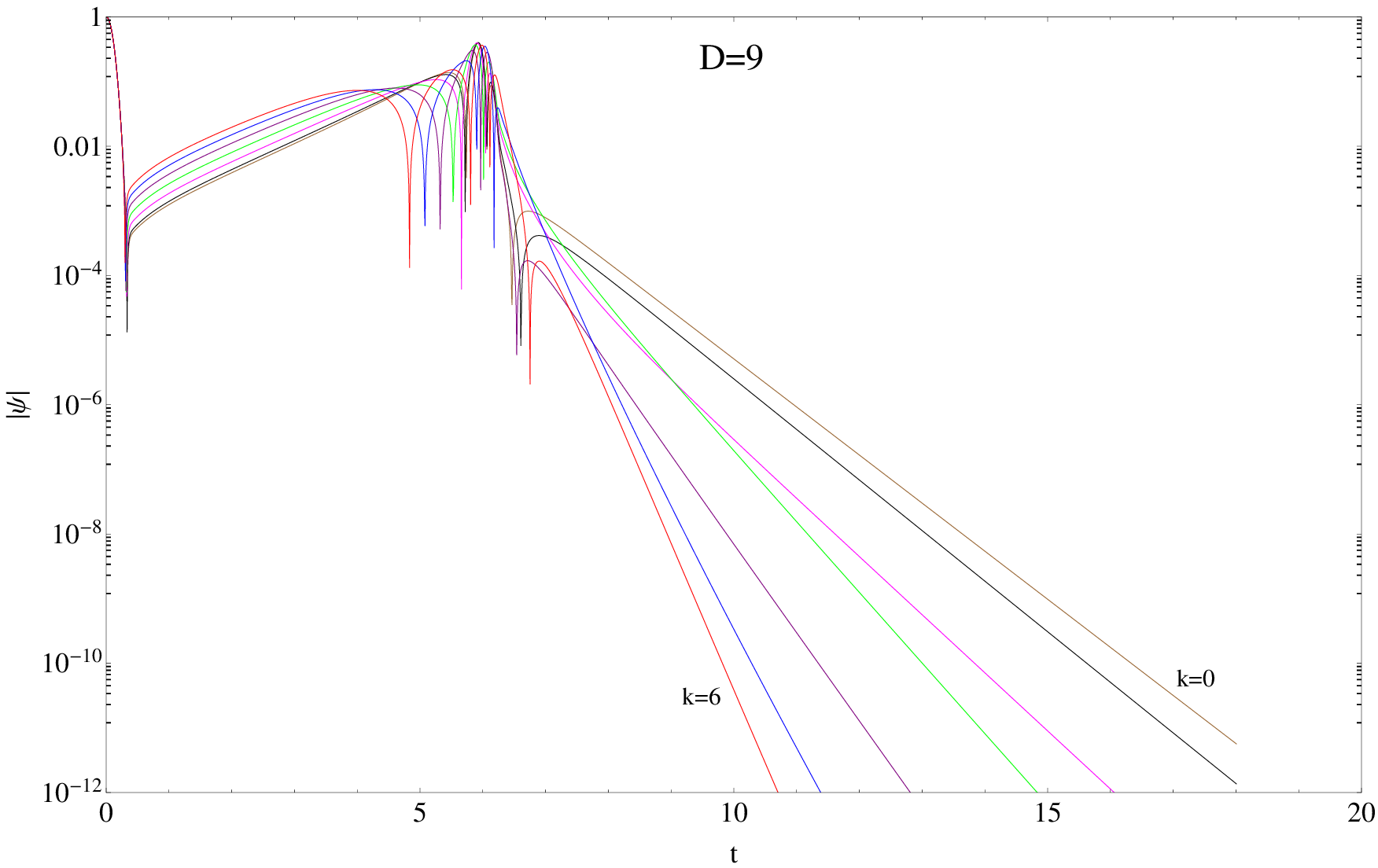}}
\vspace{0.22cm}
\scalebox{0.27}{\includegraphics{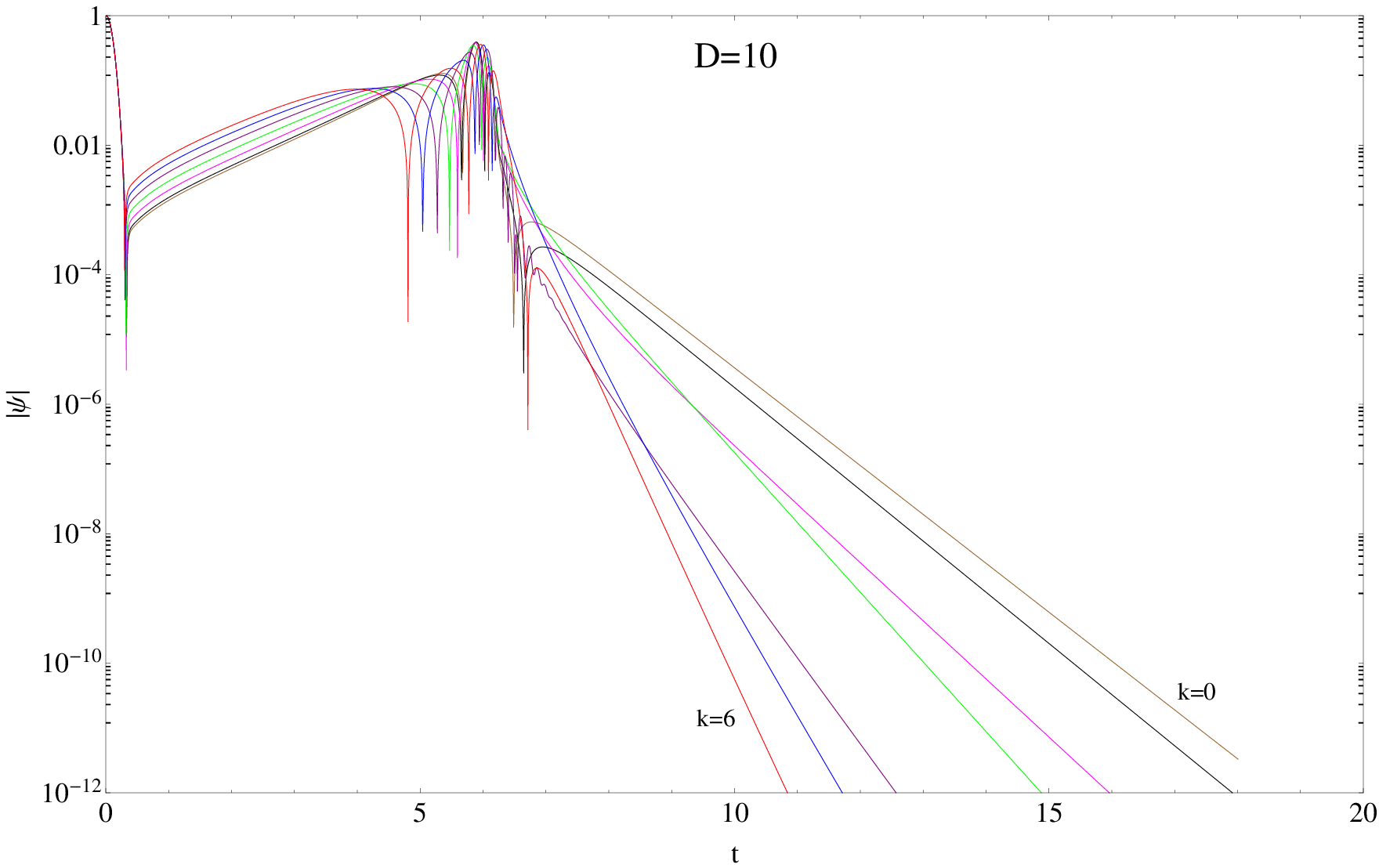}}
\end{center}
\caption{\it Semi-logarithmic plots of the time-domain evolution of scalar perturbations in Lifshitz black holes for various dimensions,
for $M=L=m=1$ and from top to bottom curves, $k=0,...6$.} \label{perfiles2}
\end{figure}
\begin{figure}[t]
\begin{center}
\scalebox{0.25}{\includegraphics{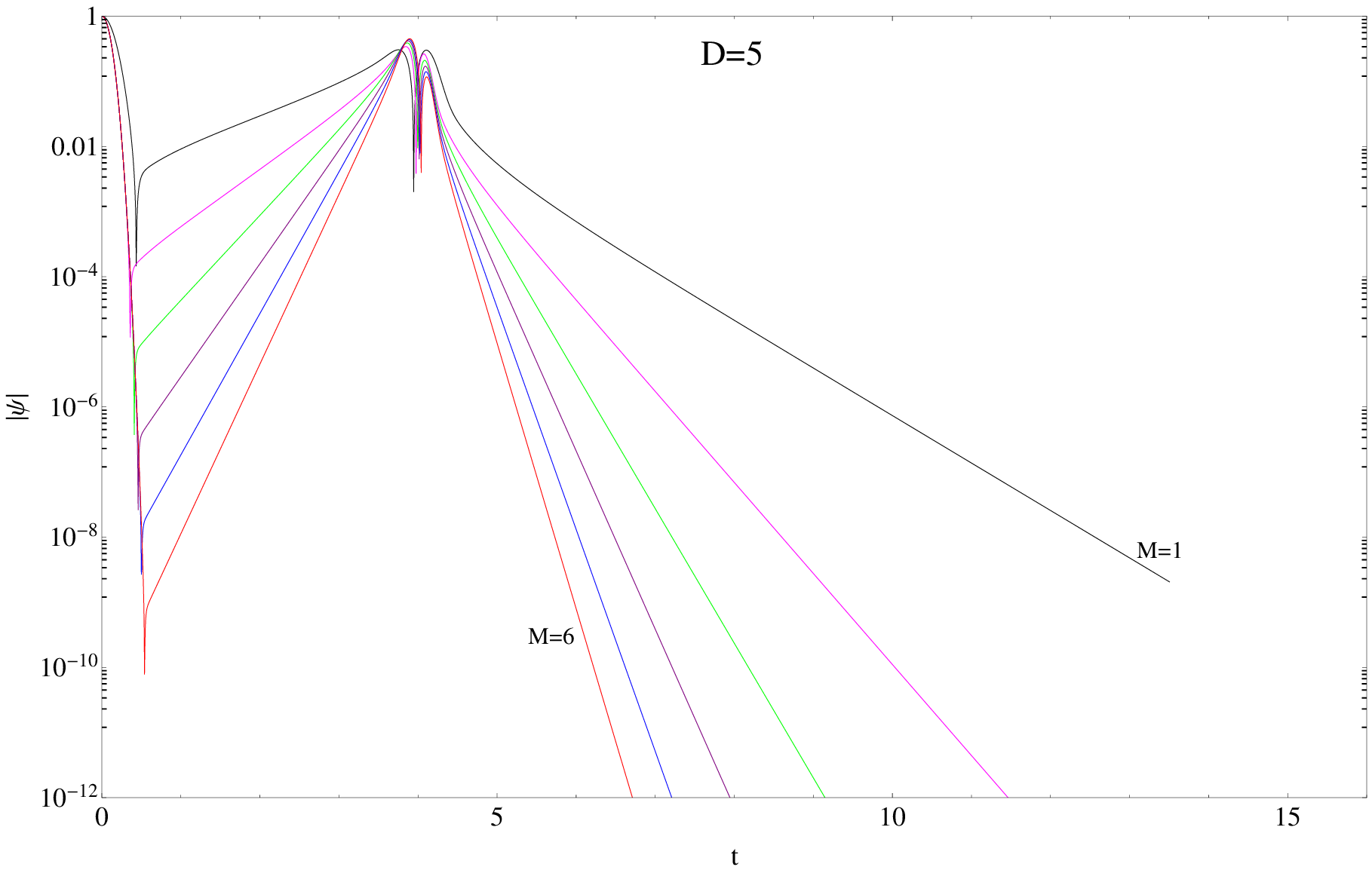}}
\vspace{0.22cm}
\scalebox{0.25}{\includegraphics{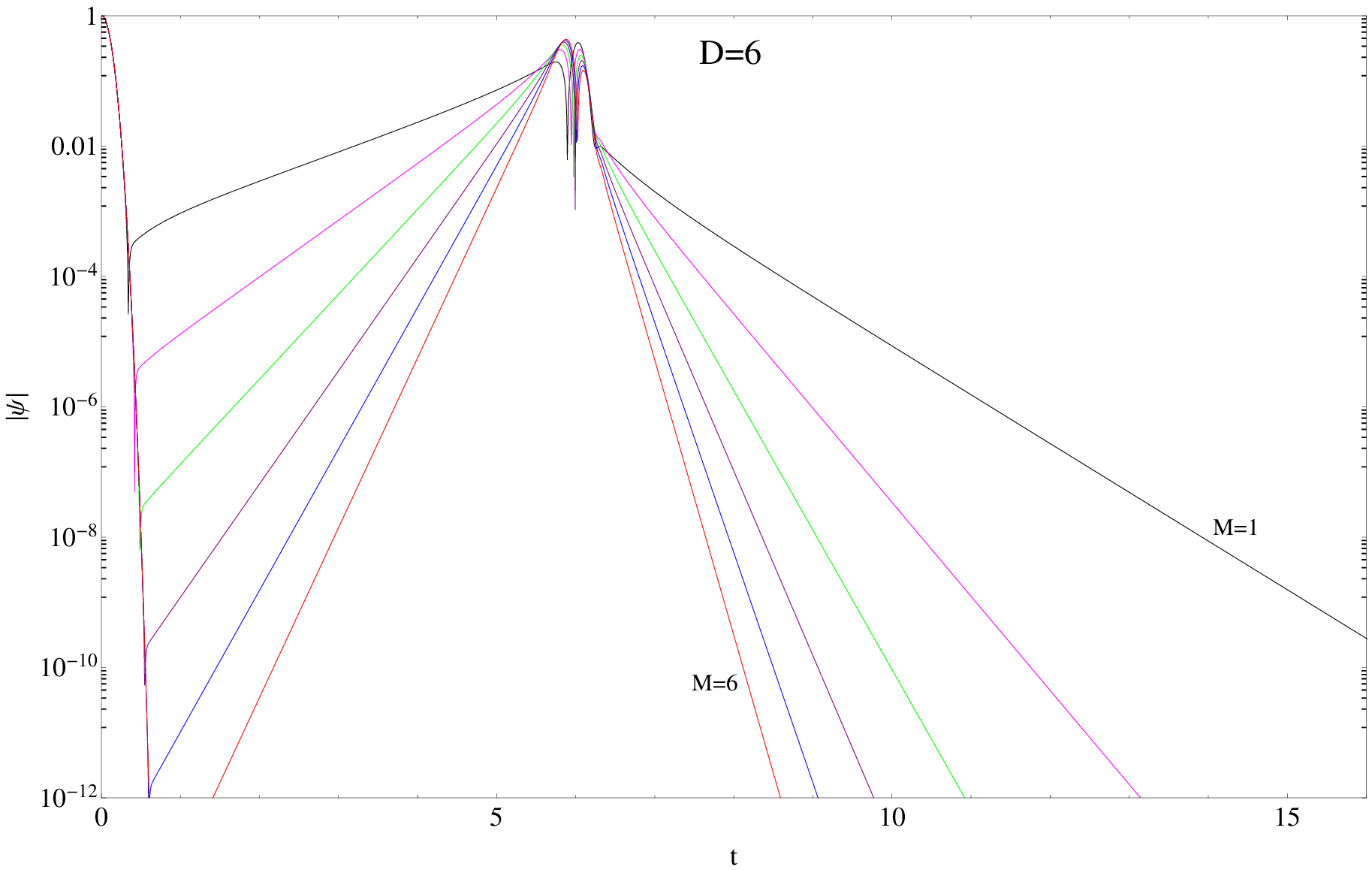}}
\vspace{0.22cm}
\scalebox{0.25}{\includegraphics{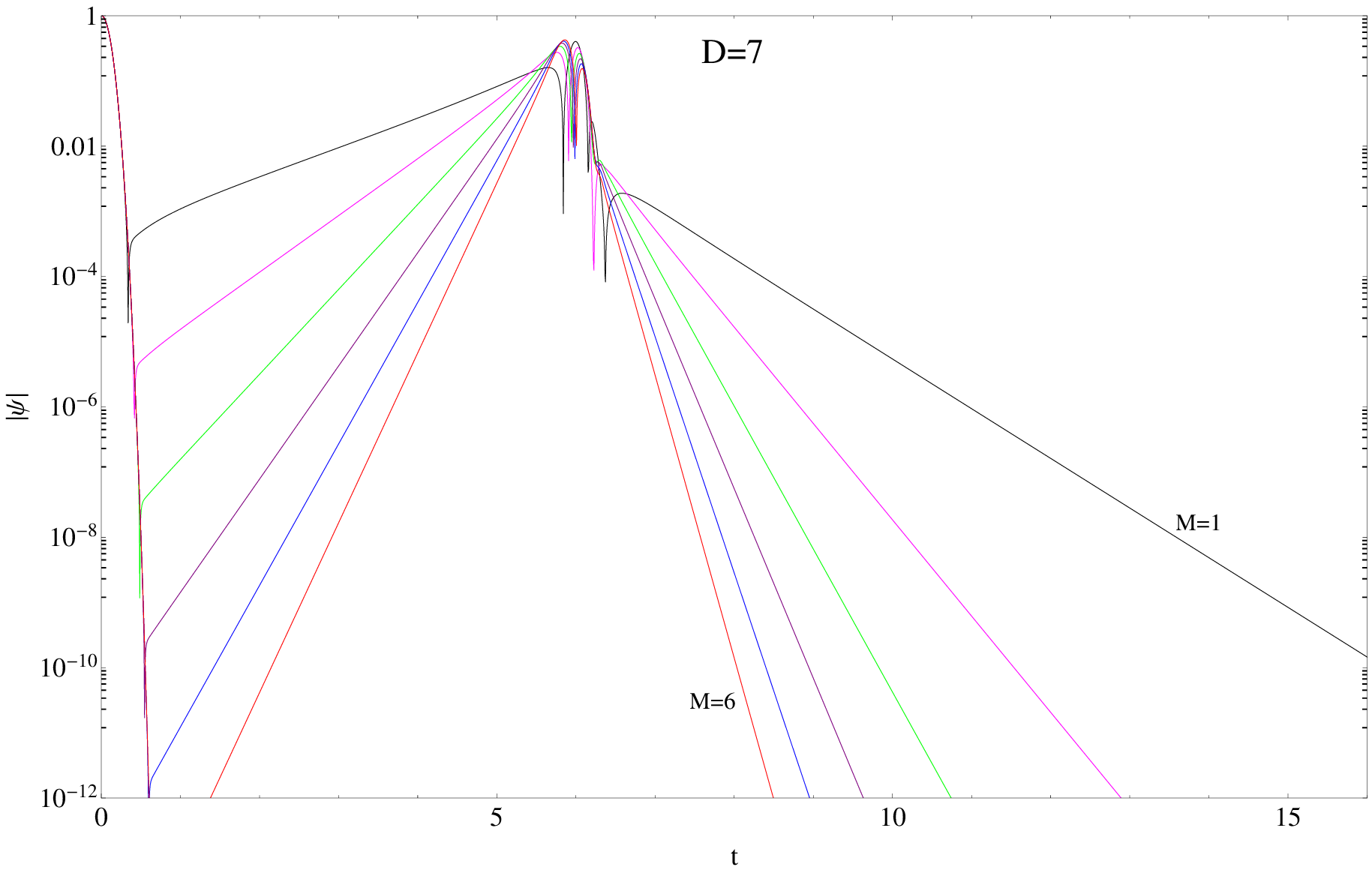}}
\vspace{0.22cm}
\scalebox{0.25}{\includegraphics{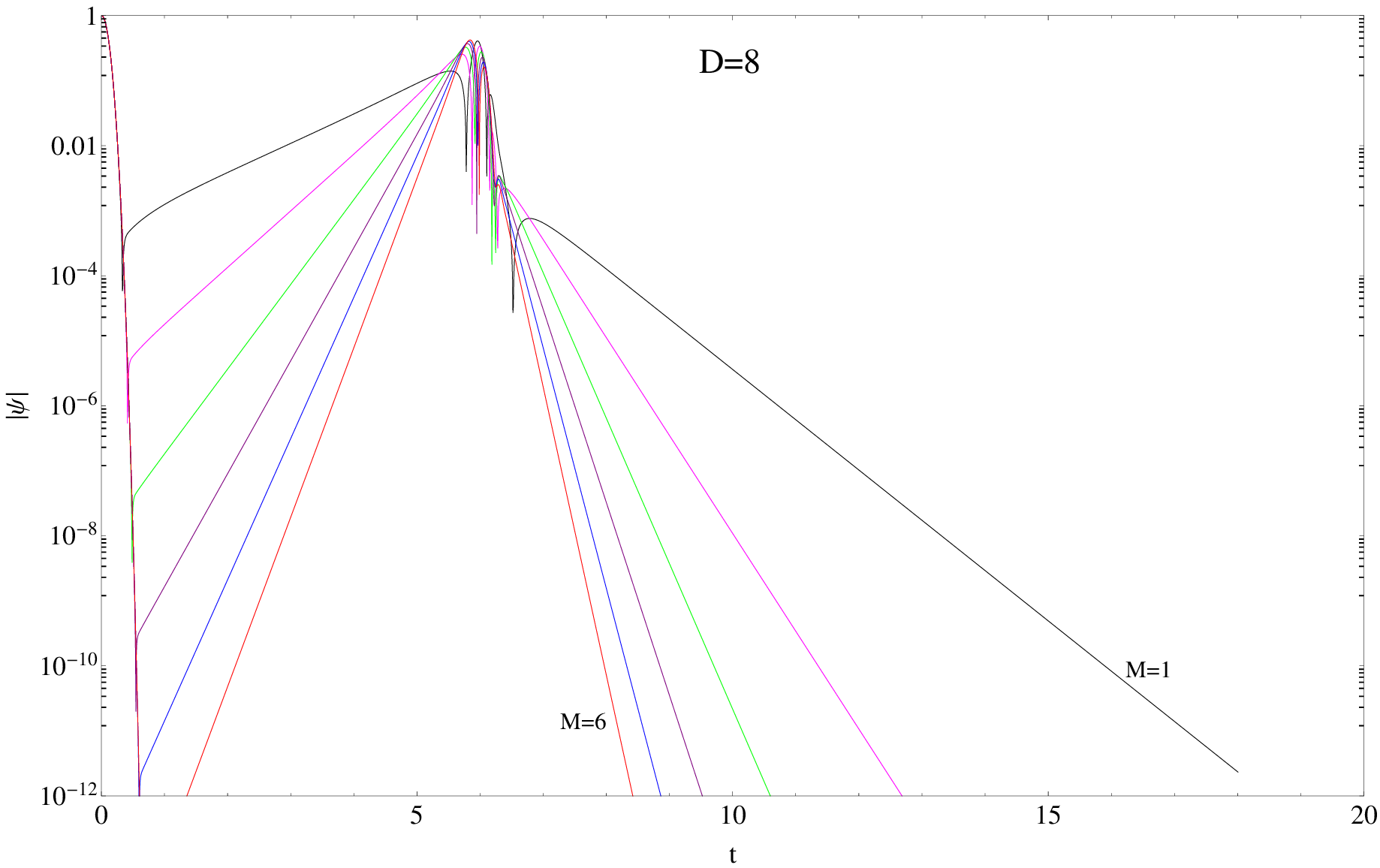}}
\vspace{0.22cm}
\scalebox{0.27}{\includegraphics{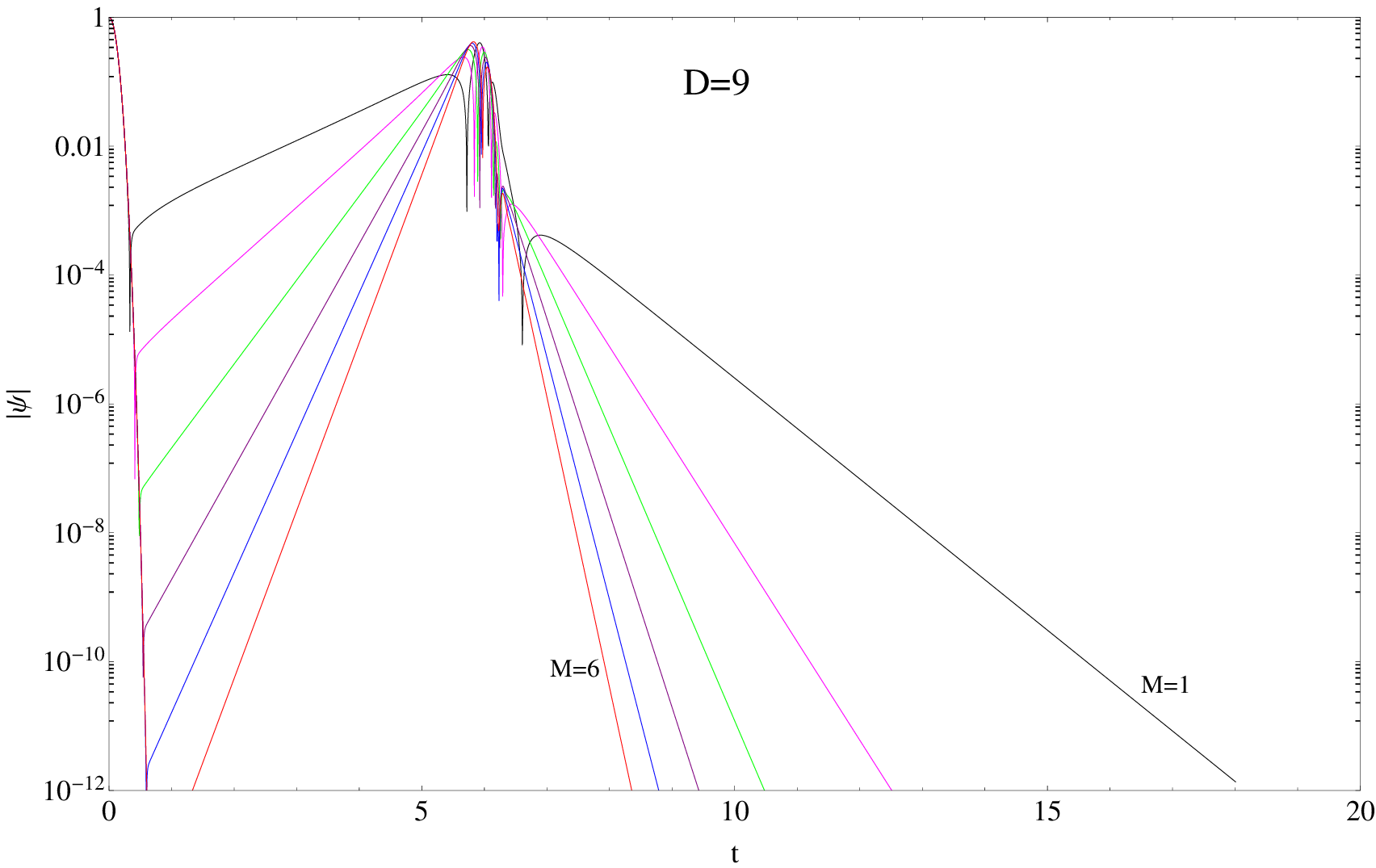}}
\vspace{0.22cm}
\scalebox{0.27}{\includegraphics{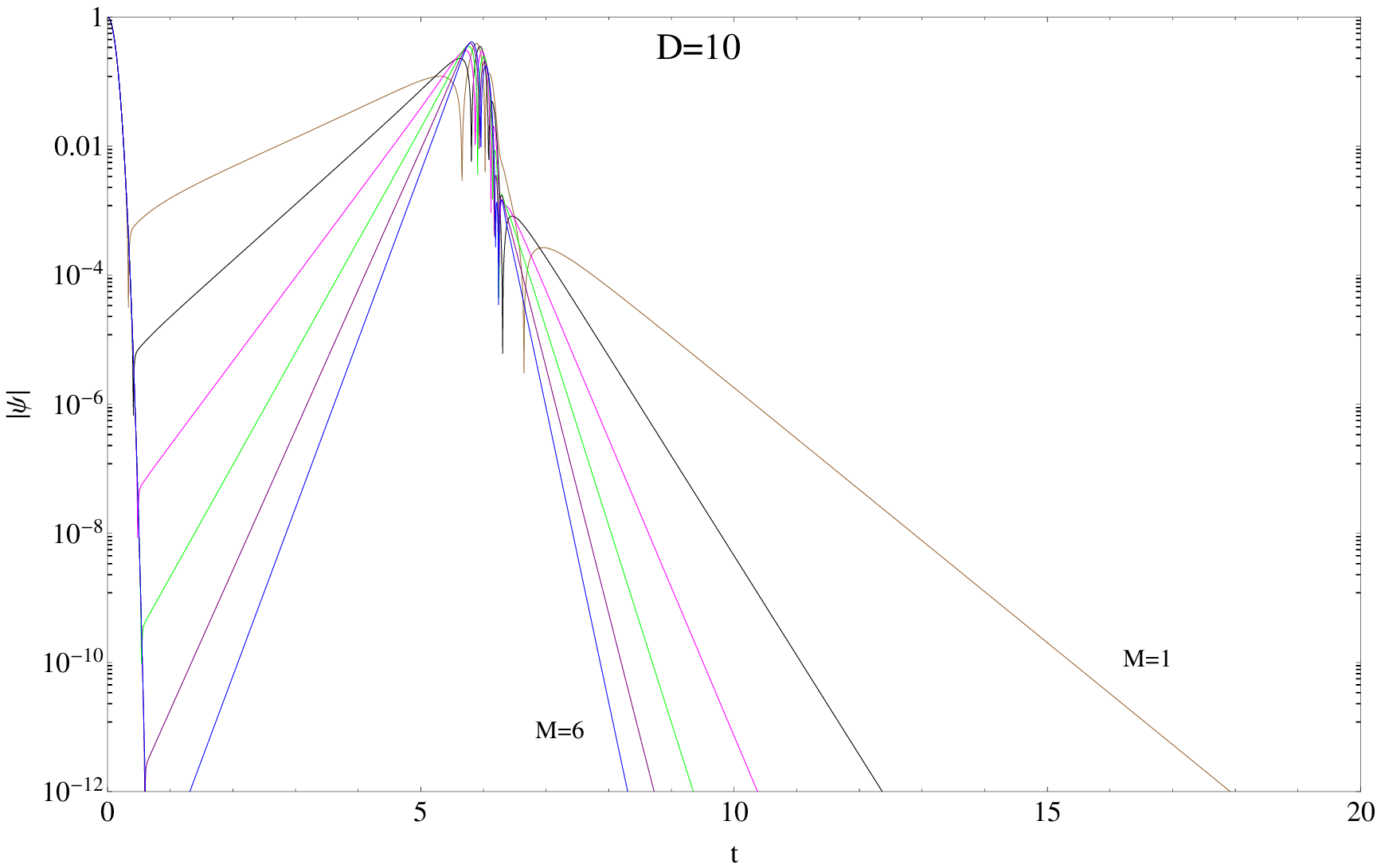}}
\end{center}
\caption{\it Semi-logarithmic plots of the time-domain evolution of scalar perturbations in Lifshitz black holes for
various dimensions, for $k=L=m=1$ and from top to bottom curves, $M=1,...6$.} \label{perfiles3}
\end{figure}
\begin{figure}[t]
\begin{center}
\scalebox{0.27}{\includegraphics{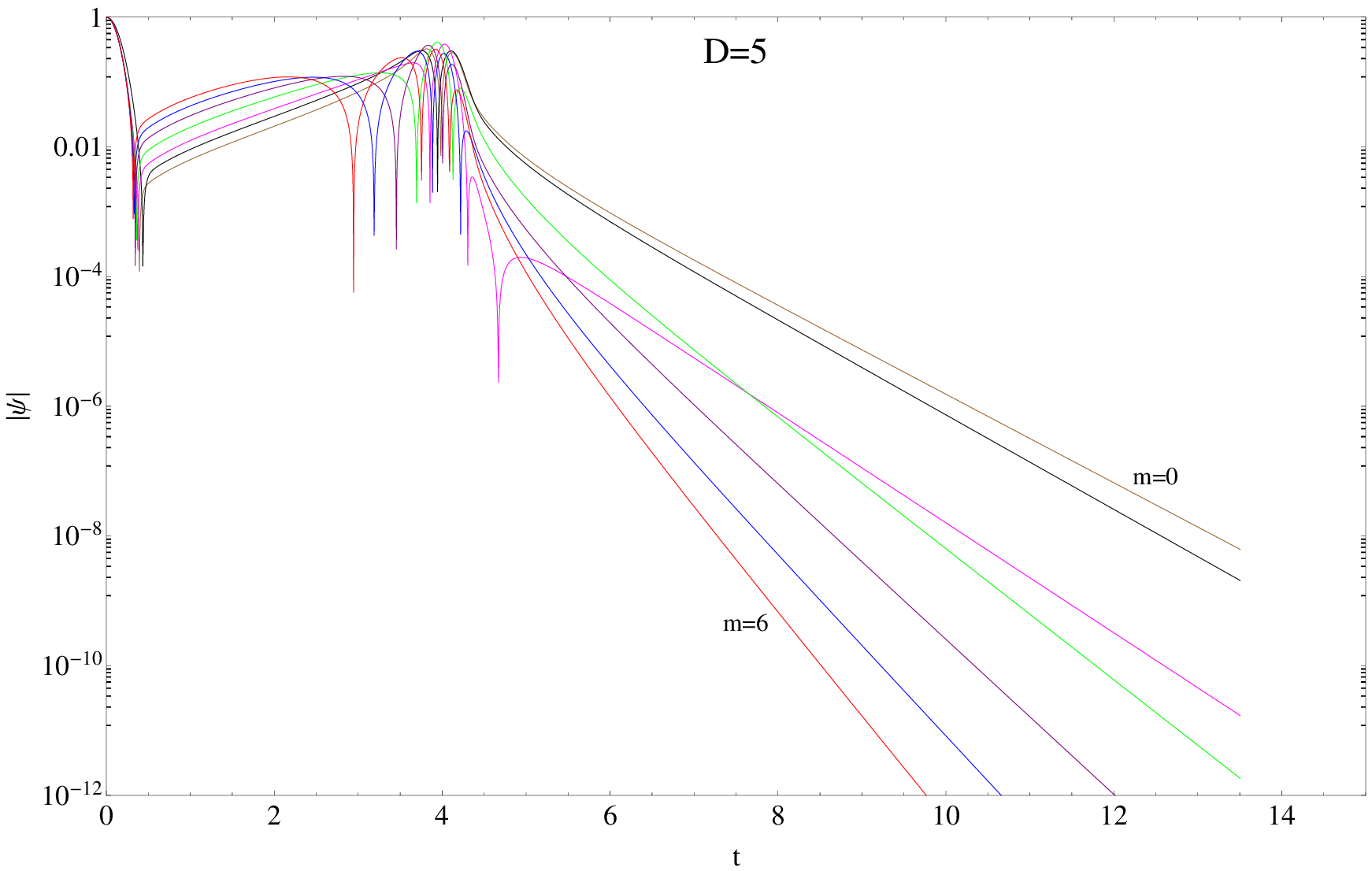}}
\vspace{0.22cm}
\scalebox{0.27}{\includegraphics{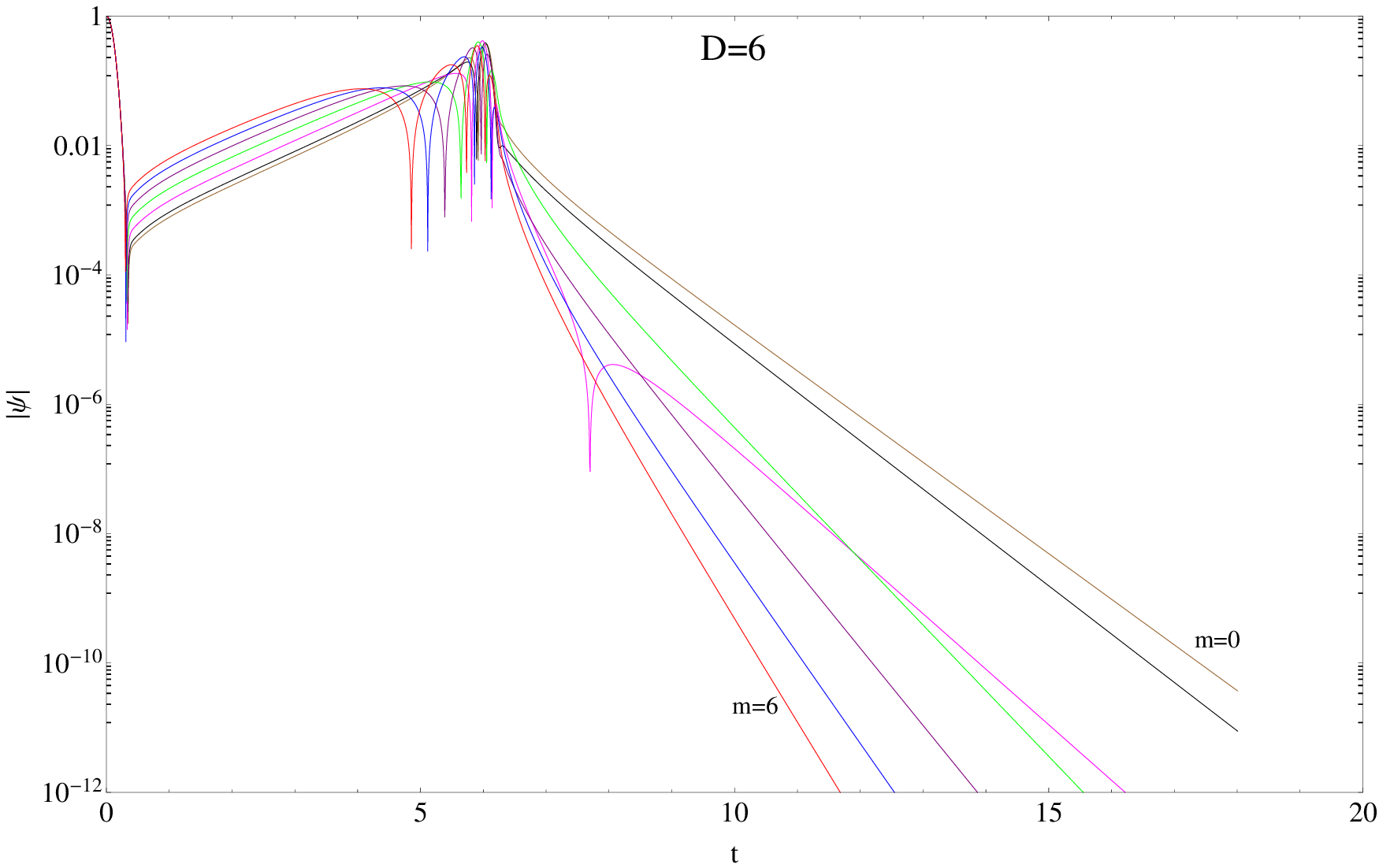}}
\vspace{0.22cm}
\scalebox{0.27}{\includegraphics{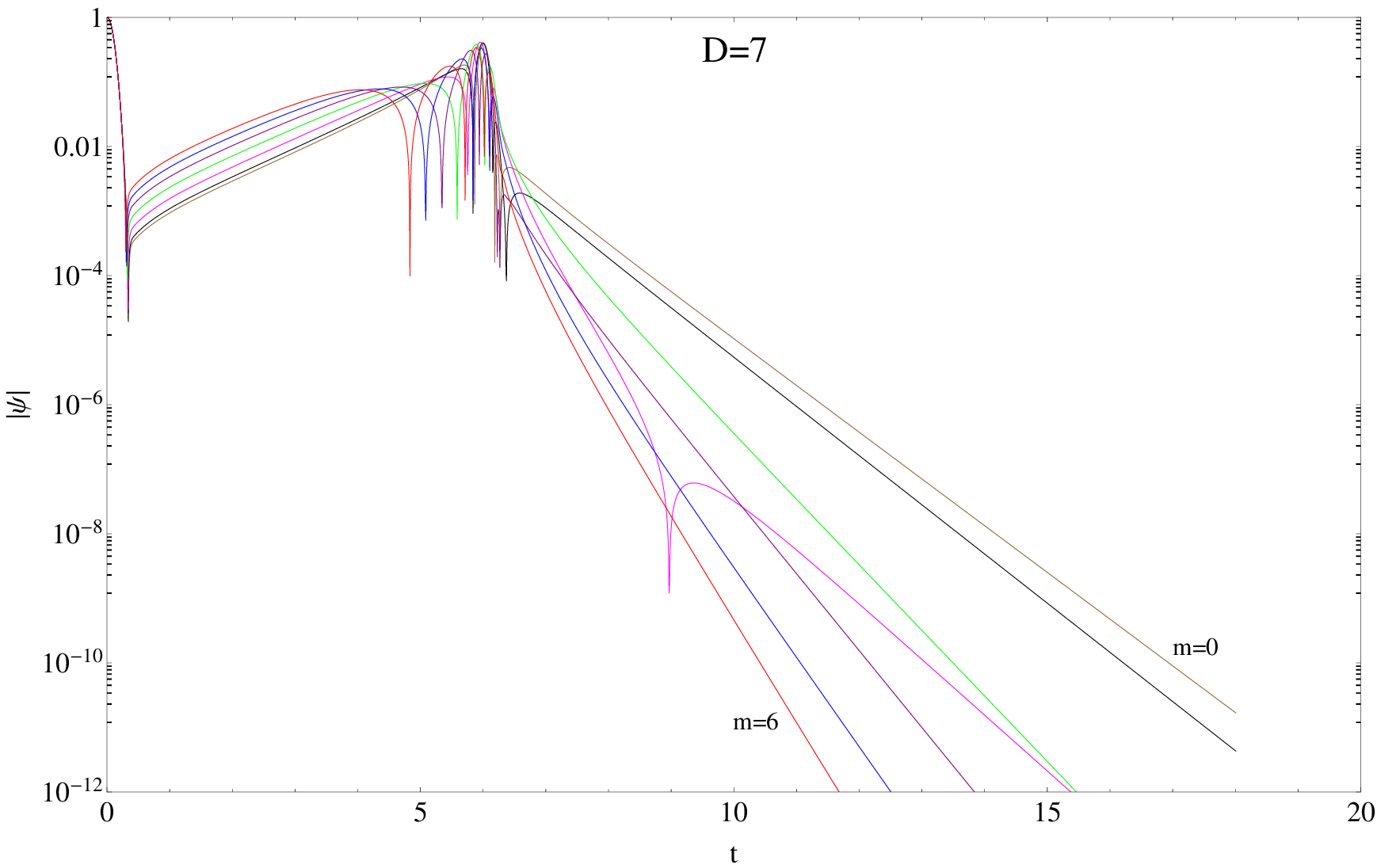}}
\vspace{0.22cm}
\scalebox{0.27}{\includegraphics{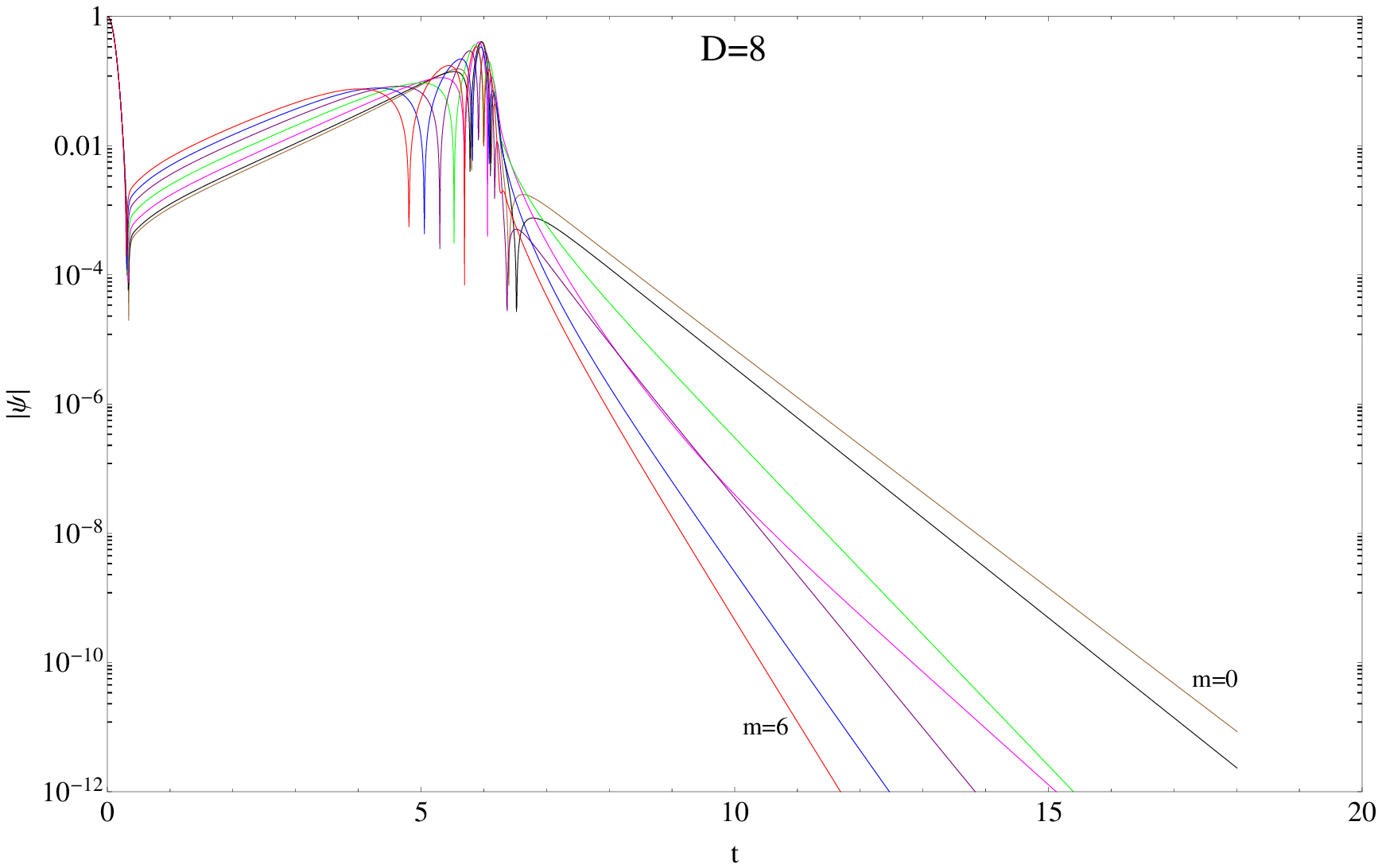}}
\vspace{0.22cm}
\scalebox{0.27}{\includegraphics{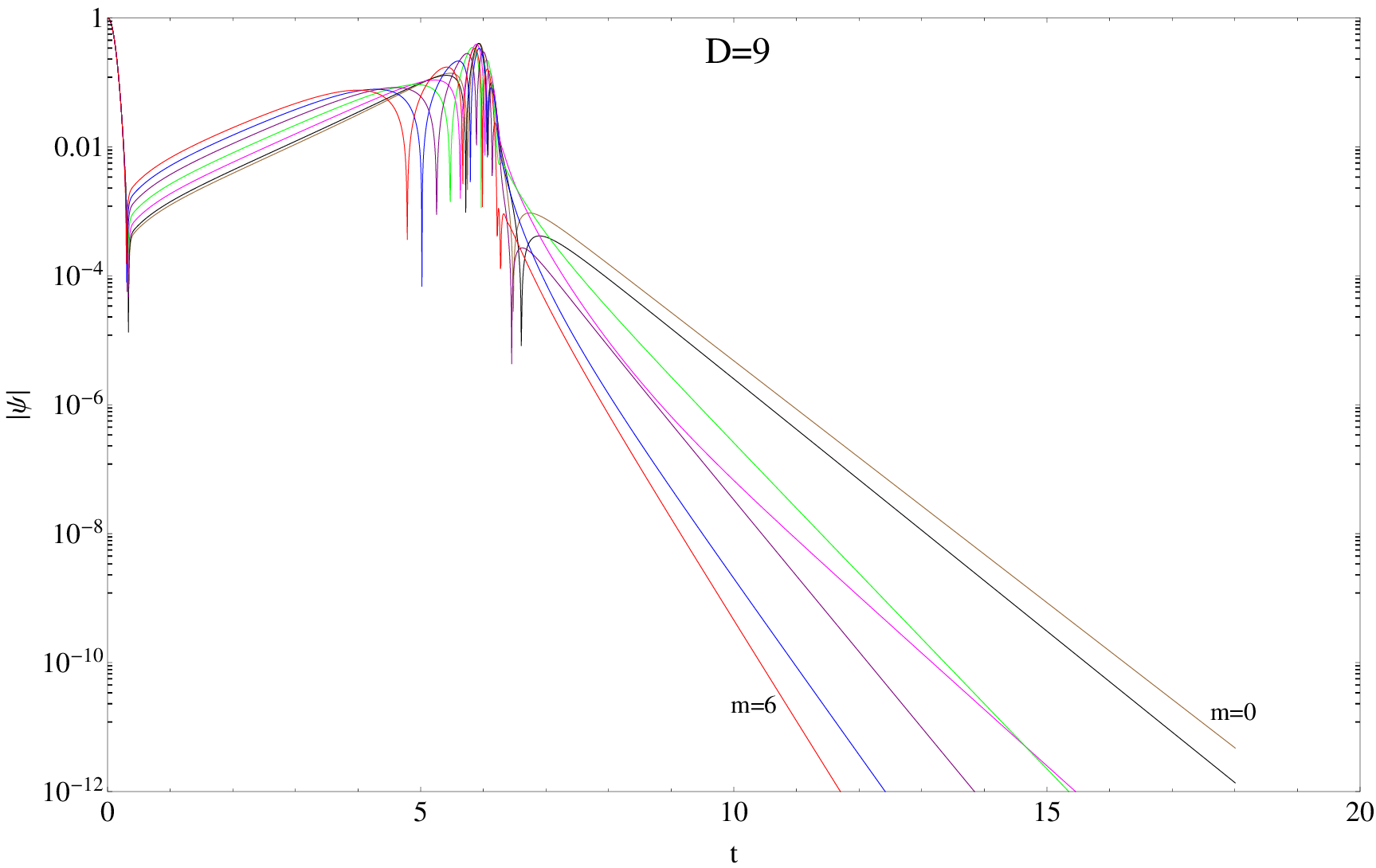}}
\vspace{0.22cm}
\scalebox{0.27}{\includegraphics{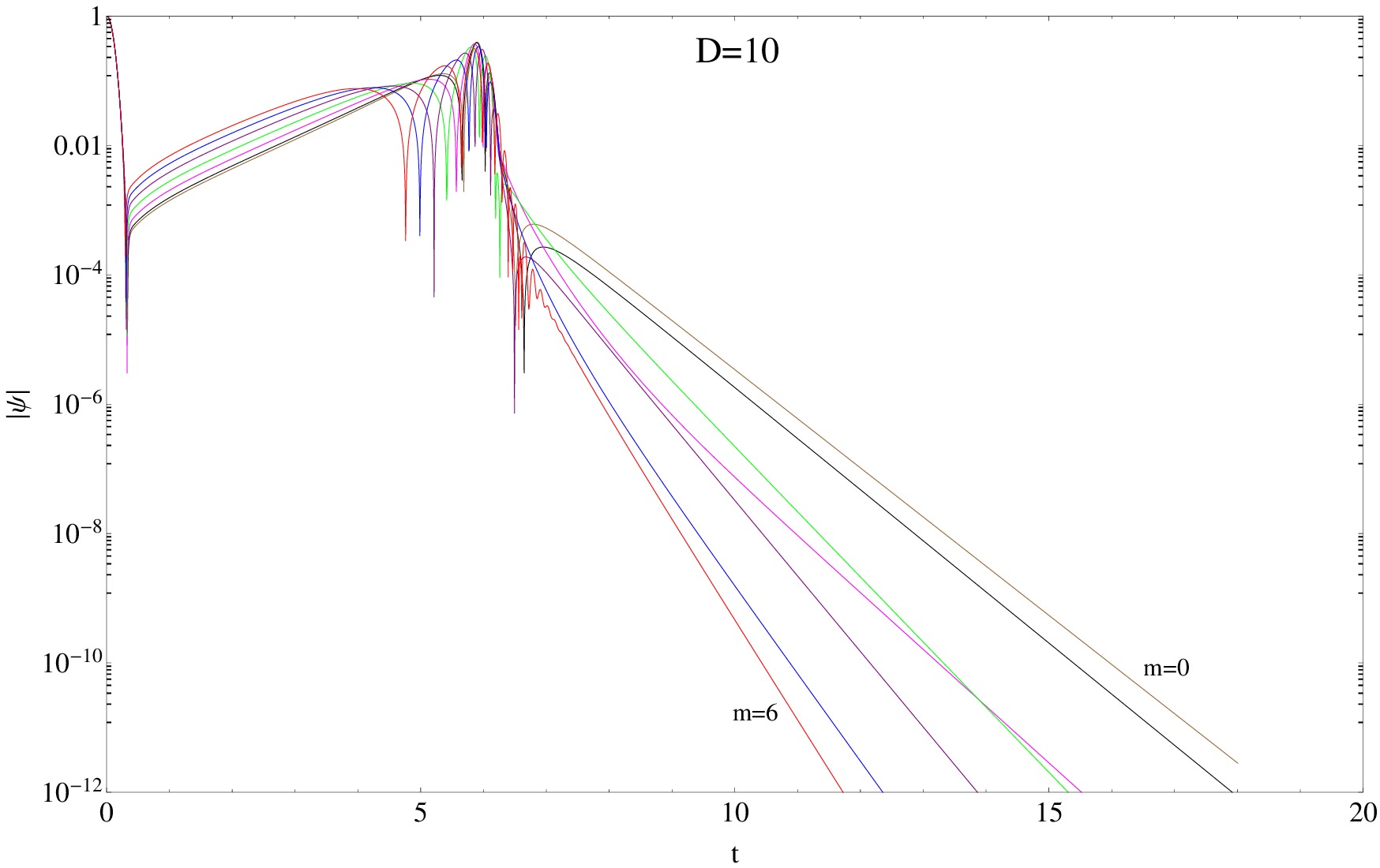}}
\end{center}
\caption{\it Semi-logarithmic plots of the time-domain evolution of scalar perturbations in Lifshitz black holes for various dimensions,
for $M=L=k=1$ and from top to bottom curves, $m=0,...6$.} \label{perfiles4}
\end{figure}
Equation (\ref{timeevolescalar}) can be solved numerically by finite difference method. Taking $t = t_{0} + k \Delta t $ and $r_{*}= r_{*0}+j
\Delta r_{*} $, substituting in (\ref{timeevolescalar}) and rearranging the terms we obtain the difference equation
\begin{equation}\label{edf1}
\psi_{k+1}^{j}=-\psi_{k-1}^{j}+\frac{\Delta t^{2}}{\Delta r_{*}^{2}}\left(\psi_{k}^{j+1}+\psi_{k}^{j-1} \right)+\left(2-
2\frac{\Delta t^{2}}{\Delta r_{*}^{2}}- V_{j}\right)\psi_{k}^{j}
\end{equation}
As initial conditions we choose the static gaussian
\begin{equation}\label{edf1}
\psi \left(r_{*},t_{0}\right)=Ae^{-a(r_{*}-b)^{2}}, \ \ \ \ \ \frac{\partial}{\partial t}\psi\left(r_{*},t\right)|_{t=t_{0}}=0
\end{equation}
We also impose Dirichlet conditions at the AdS boundary $\psi (r_{*}=0,t)=0$.
The Von Neumann stability condition applied to the above difference equations results in the relation
\[
\frac{\Delta t^{2}}{\Delta r_{*}{^2}}\sin^{2}\left(\frac{\alpha }{2}\right)+\frac{\Delta t^{2}}{4}V(\alpha)<1
\]
Then, defining $V_{\max}$ as the largest value of $V_{j}$ in the numerical grid, our numerical solution will be stable if
\begin{equation}\label{VNstab}
\frac{\Delta t^{2}}{\Delta r_{*}^{2}}+\frac{\Delta t^{2}}{4}V_{\max }<1
\end{equation}
Now taking $\Delta t=\frac{1}{2}\Delta r_{*}$ and choosing a sufficiently small time step such that $V_{max}<\frac{3}{\Delta t^{2}}$ we
assure the stability of our numerical codes. In all the calculations performed by us we verified the fulfillment of this stability
condition.

As Figures \ref{perfiles1}-\ref{perfiles4} shows, the time evolution of scalar fluctuations in $z=2$ Lifshitz black holes can be divided
in two stages. The first one is a transient response of the field due to its initial interaction with the black hole, and is
strongly dependent on the initial conditions for the perturbation. This transient stage is followed by purely damped quasinormal
modes, without oscillations, in contrast with the usual behavior of asymptotically AdS black holes, where quasinormal oscillations
caracterized by frequencies with well defined real and imaginary parts dominate all the evolution after the transient phase.

The above picture is valid for all dimensions, and values of the parameters $M$, $L$, $m$ and $k$ that characterized the background space-time
and the scalar field fluctuations. As the dimensionality of the space-times increases, the amplitude of massless and massive scalar
fluctuations decays more quickly. Then, Lifshitz black holes becomes more rigid for higher dimensions. The same behavior is observed
for scalar fluctuations with increasing wave-number $k$, for all dimensions. For fixed dimensionality of the background, $s$-wave
fluctuations are more longer lived. For fixed $k$ the higher the dimension of the space-time, the lower the decay rate of the perturbation.

When the mass of black hole $M$  and that of the scalar field fluctuation $m$ change, we observe, for $D$ and $k$ fixed, an
increase of the decay rate of the perturbation with the increase of those parameters. Then, the propagation of scalar perturbations
outside higher dimensional Lifshitz black hole exhibits a very simple behavior.

All of the above results are confirmed after the calculation of the quasinormal frequencies of the scalar perturbations in the systems
considered. The quasinormal modes are solutions of equations (\ref{exact1}) or (\ref{timeevolescalar})
with the specific boundary conditions requiring purely outgoing waves
at spatial infinity and purely incoming waves on the event horizon $r_{+}\equiv \sqrt{M}L$.

In practical numerical calculations, the time evolution data obtained solving the evolution equations mix the fundamental quasinormal mode
and the higher overtones. As higher overtones decay more quickly that the fundamental mode, a common recipe is to fit the numerical data far
away from the beginning of the quasinormal ringing, to be sure that higher overtones are suppressed.

Another numerical method to determine the quasinormal frequencies for AdS space-times, that can be easily adapted to asymptotically
Lifshitz space-times is the Horowitz-Hubbeny method \cite{horowitz-hubeny}.

Tables I to VI show the numerical results obtained for the quasinormal frequencies of higher dimensional Lifshitz black holes with dynamical exponent $z=2$, using the above mentioned numerical methods. We also include the exact results that can be obtained solving analytically the evolution equation, as we will show in the next section.

As we can observe from the tables, there exist a very good coincidence between the analytical results and the numerical ones. The
Horowitz-Hubbeny method appears to be the best to perform numerical calculations and the time domain integrations lead to little
worse results, due to the presence of numerical errors and precision problems in a practical calculation. However, if an analytical
formula is absent, the Horowitz-Hubbeny method is very difficult to implement, because we need an approximate knowledge of the
quasinormal frequency to complete the calculation, a fact that indicates the importance of develop time domain integration.

It is quite surprising that the evolution of perturbations is extremely simple. Indeed, the decay is basically that
of an exponentially decaying tail, which means no real part of the frequency. This sounds like no quasi normal
decay. This may be conjectured as a normal phase, or no superconductor at the border. In figure (\ref{perfiles1})
one sees that the perturbations boringly die out leaving no sound of the Black Hole, just showing that perturbations
spread and disappear. This result is independent of the dimensionality of space time and spin number. Therefore, the
structure of Lifshitz Black Hole does not seem to lead to any interesting critical phenomena at the border.

Moreover, the incredible similarity of decays, almost independently of dimension or $m$-value, with strikingly small differences
in the values of the constants, shows that the $z=2$ Black Holes are much more stable than the Einstein gravity counterparts.
This leads again to the conjecture that  the CFT dual never undergoes phase transitions. Actually, this result
goes in the same line as the three-dimensional New Massive Gravity, whose CFT dual is conjectured to be an integrable model,
namely, KdV equation \cite{abdallaoliveira}.

\begin{center}
\begin{table}[htb!]
   \setlength{\belowcaptionskip}{5pt}  
  {\begin{tabular}{|c|c|c|c|c|c|}
      \hline
      \hline
       \multicolumn{1}{|c|}{\ \ $M$ \ \ }& \multicolumn{1}{|c|}{\ \ $k$ \ \ }& \multicolumn{1}{|c|}{\ \ $m$ \ \ }&
\multicolumn{1}{|c|}{\ \ Time integration \ \ }&\multicolumn{1}{|c|}{\ \ Horowitz-Hubbeny \ \ }&\multicolumn{1}{|c|}{ \ \ Analytical \ \ } \\
      \hline
      $1$ & $1$ & $1$ & $-1.677057i$ & $-1.677033i$&$ \ \ -1.677033i \ \ $ \\
      \hline
  $2$ & $1$ & $1$ & $-3.218724i$ & $-3.218659i$&$ -3.218659i$ \\
      \hline
  $3$ & $1$ & $1$ & $-4.760508i$ & $-4.760286i$&$ -4.760286i$ \\
      \hline
      $4$ & $1$ & $1$ & $-6.302442i$ & $-6.301912i$&$-6.301912i$ \\
     \hline
     $5$ & $1$ & $1$ & $-7.844577i$ & $-7.843538i$&$ -7.843538i$ \\
     \hline
$1$ & $0$ & $1$ & $-1.541628i$ & $-1.541626i$&$ -1.541626i$ \\
      \hline
$1$ & $2$ & $1$ & $-2.083253i$ & $-2.083253i$&$ -2.083253i$ \\
      \hline
$1$ & $3$ & $1$ & $-2.760615i$ & $-2.760286i$&$ -2.760286i$ \\
     \hline
       $1$ & $4$ & $1$ & $-3.709838i$ & $-3.708132i$&$ -3.708132i$ \\
      \hline
 $1$ & $5$ & $1$ & $-4.917540i$ & $-4.926791i$&$ -4.926791i$ \\
      \hline
     $1$ & $1$ & $0$ & $-1.571455i$ & $-1.571429i$&$-1.571429i$ \\
      \hline
      $1$ & $1$ & $2$ & $-1.952045i$ & $-1.952027i$&$ -1.952027i$ \\
      \hline
 $1$ & $1$ & $3$ & $-2.325160i$ & $-2.325145i$&$ -2.325145i$ \\
     \hline
  $1$ & $1$ & $4$ & $-2.749181i$ & $-2.749172i$&$ -2.749172i$ \\
     \hline
     $1$ & $1$ & $5$ & $-3.200247i$ & $-3.200247i$&$ -3.200247i$ \\
          \hline
       \hline
  \end{tabular}\label{cinco}}
   \caption{\it Scalar quasinormal frequencies in $5$-dimensional $L=1$ Lifshitz black holes for various values of $M$, $m$ and $k$.
Results of calculation using time domain integration, Horowitz-Hubbeny method and exact analytical formula are included.}
      \end{table}
      \end{center}
\begin{center}
\begin{table}[htb!]
   \setlength{\belowcaptionskip}{5pt}  
  {\begin{tabular}{|c|c|c|c|c|c|}
      \hline
      \hline
       \multicolumn{1}{|c|}{\ \ $M$ \ \ }& \multicolumn{1}{|c|}{\ \ $k$ \ \ }& \multicolumn{1}{|c|}{\ \ $m$ \ \ }&
\multicolumn{1}{|c|}{\ \ Time integration \ \ }&\multicolumn{1}{|c|}{\ \ Horowitz-Hubbeny \ \ }&\multicolumn{1}{|c|}{ \ \ Analytical \ \ } \\
      \hline
      $1$ & $1$ & $1$ & $-1.720780i$ & $-1.720759i$&$ \ \ -1.720759i \ \ $ \\
      \hline
  $2$ & $1$ & $1$ & $-3.321490i$ & $-3.321392i$&$ -3.321392i$ \\
      \hline
  $3$ & $1$ & $1$ & $-4.922364i$ & $-4.922025i$&$ -4.922025i$ \\
      \hline
      $4$ & $1$ & $1$ & $-6.523469i$ & $ -6.522657i$&$ -6.522657i$ \\
     \hline
     $5$ & $1$ & $1$ & $-8.124882i$ & $-8.123290i$&$ -8.123290i$ \\
     \hline
$1$ & $0$ & $1$ & $-1.600632i$ & $-1.600633i$&$ -1.600633i$ \\
      \hline
$1$ & $2$ & $1$ & $-2.081199i$ & $-2.081139i$&$ -2.081139i$ \\
      \hline
$1$ & $3$ & $1$ & $-2.682273i$ & $-2.681771i$&$ -2.681771i$ \\
     \hline
       $1$ & $4$ & $1$ & $-3.523343i$ & $-3.522657i$&$ -3.522657i$ \\
      \hline
 $1$ & $5$ & $1$ & $-4.565130i$ & $-4.603787i$&$-4.603787i$ \\
      \hline
     $1$ & $1$ & $0$ & $-1.571455i$ & $-1.571455i$&$-1.571455i$ \\
      \hline
      $1$ & $1$ & $2$ & $-1.977047i$ & $-1.977082i$&$ -1.977082i$ \\
      \hline
 $1$ & $1$ & $3$ & $-2.335233i$ & $-2.335205i$&$ -2.335205i$ \\
     \hline
  $1$ & $1$ & $4$ & $-2.750061i$ & $-2.750000i$&$ -2.750000i$ \\
     \hline
     $1$ & $1$ & $5$ & $-3.659501i$ & $-3.195887i$&$ -3.195887i$ \\
          \hline
       \hline
  \end{tabular}\label{seis}}
   \caption{\it Scalar quasinormal frequencies in $6$-dimensional $L=1$ Lifshitz black holes for various values of $M$, $m$ and $k$.
Results of calculation using time domain integration, Horowitz-Hubbeny method and exact analytical formula are included.}
      \end{table}
      \end{center}
\begin{center}
\begin{table}[htb!]
   \setlength{\belowcaptionskip}{5pt}  
  {\begin{tabular}{|c|c|c|c|c|c|}
      \hline
      \hline
       \multicolumn{1}{|c|}{\ \ $M$ \ \ }& \multicolumn{1}{|c|}{\ \ $k$ \ \ }& \multicolumn{1}{|c|}{\ \ $m$ \ \ }&
\multicolumn{1}{|c|}{\ \ Time integration \ \ }&\multicolumn{1}{|c|}{\ \ Horowitz-Hubbeny \ \ }&\multicolumn{1}{|c|}{ \ \ Analytical \ \ } \\
      \hline
      $1$ & $1$ & $1$ & $-1.754315i$ & $-1.754301i$&$ \ \ -1.754301i \ \ $ \\
      \hline
  $2$ & $1$ & $1$ & $-3.400939i$ & $-3.400845i$&$ -3.400845i$ \\
      \hline
  $3$ & $1$ & $1$ & $-5.047710i$ & $-5.047389i$&$ -5.047389i$ \\
      \hline
      $4$ & $1$ & $1$ & $-6.694699i$ & $ -6.693934i$&$ -6.693934i$ \\
     \hline
     $5$ & $1$ & $1$ & $-8.341979i$ & $-8.340478i$&$ -8.340478i$ \\
     \hline
$1$ & $0$ & $1$ & $-1.646544i$ & $-1.646544i$&$ -1.646544i$ \\
      \hline
$1$ & $2$ & $1$ & $-2.077628i$ & $-2.077573i$&$ -2.077573i$ \\
      \hline
$1$ & $3$ & $1$ & $-2.616515i$ & $-2.616360i$&$ -2.616360i$ \\
     \hline
       $1$ & $4$ & $1$ & $-3.370879i$ & $-3.370661i$&$ -3.370661i$ \\
      \hline
 $1$ & $5$ & $1$ & $-4.325313i$ & $-4.340478i$&$-4.340478i$ \\
      \hline
     $1$ & $1$ & $0$ & $-1.666672i$ & $-1.666667i$&$-1.666667i$ \\
      \hline
      $1$ & $1$ & $2$ & $-1.993564i$ & $-1.993813i$&$ -1.993813i$ \\
      \hline
 $1$ & $1$ & $3$ & $-2.336972i$ & $-2.336953i$&$ -2.336953i$ \\
     \hline
  $1$ & $1$ & $4$ & $-2.741867i$ & $-2.741864i$&$ -2.741864i$ \\
     \hline
     $1$ & $1$ & $5$ & $-3.182104i$ & $-3.182091i$&$ -3.182091i$ \\
          \hline
       \hline
  \end{tabular}\label{siete}}
   \caption{\it Scalar quasinormal frequencies in $7$-dimensional $L=1$ Lifshitz black holes for various values of $M$, $m$ and $k$.
Results of calculation using time domain integration, Horowitz-Hubbeny method and exact analytical formula are included.}
      \end{table}
      \end{center}
\begin{center}
\begin{table}[htb!]
   \setlength{\belowcaptionskip}{5pt}  
  {\begin{tabular}{|c|c|c|c|c|c|}
      \hline
      \hline
       \multicolumn{1}{|c|}{\ \ $M$ \ \ }& \multicolumn{1}{|c|}{\ \ $k$ \ \ }& \multicolumn{1}{|c|}{\ \ $m$ \ \ }&
\multicolumn{1}{|c|}{\ \ Time integration \ \ }&\multicolumn{1}{|c|}{\ \ Horowitz-Hubbeny \ \ }&\multicolumn{1}{|c|}{ \ \ Analytical \ \ } \\
      \hline
      $1$ & $1$ & $1$ & $-1.7807875i$ & $-1.780776i$&$ \ \ -1.780776i \ \ $ \\
      \hline
  $2$ & $1$ & $1$ & $-3.464048i$ & $-3.463956i$&$ -3.463956i$ \\
      \hline
  $3$ & $1$ & $1$ & $-5.147450i$ & $-5.147135i$&$ -5.147135i$ \\
      \hline
      $4$ & $1$ & $1$ & $-6.831065i$ & $ -6.830314i$&$ -6.830314i$ \\
     \hline
     $5$ & $1$ & $1$ & $-8.514963i$ & $-8.513494i$&$ -8.513494i$ \\
     \hline
$1$ & $0$ & $1$ & $-1.683179i$ & $-1.683179i$&$ -1.683179i$ \\
      \hline
$1$ & $2$ & $1$ & $-2.073610i$ & $-2.073567i$&$ -2.073567i$ \\
      \hline
$1$ & $3$ & $1$ & $-2.561607i$ & $-2.561553i$&$ -2.561553i$ \\
     \hline
       $1$ & $4$ & $1$ & $-3.244799i$ & $-3.244732i$&$-3.244732i$ \\
      \hline
 $1$ & $5$ & $1$ & $-4.111895i$ & $-4.123106i$&$-4.123106i$ \\
      \hline
     $1$ & $1$ & $0$ & $-1.700032i$ & $-1.700000i$&$-1.700000i$ \\
      \hline
      $1$ & $1$ & $2$ & $-2.005357i$ & $-2.005092i$&$ -2.005092i$ \\
      \hline
 $1$ & $1$ & $3$ & $-2.333347i$ & $-2.333333i$&$ -2.333333i$ \\
     \hline
  $1$ & $1$ & $4$ & $-2.727540i$ & $-2.727543i$&$ -2.727543i$ \\
     \hline
     $1$ & $1$ & $5$ & $-3.161256i$ & $-3.161250i$&$ -3.161250i$ \\
          \hline
       \hline
  \end{tabular}\label{ocho}}
   \caption{\it Scalar quasinormal frequencies in $8$-dimensional $L=1$ Lifshitz black holes for various values of
$M$, $m$ and $k$. Results of calculation using time domain integration, Horowitz-Hubbeny method and exact analytical formula are included.}
      \end{table}
      \end{center}
\begin{center}
\begin{table}[htb!]
   \setlength{\belowcaptionskip}{5pt}  
  {\begin{tabular}{|c|c|c|c|c|c|}
      \hline
      \hline
       \multicolumn{1}{|c|}{\ \ $M$ \ \ }& \multicolumn{1}{|c|}{\ \ $k$ \ \ }& \multicolumn{1}{|c|}{\ \ $m$ \ \ }&
\multicolumn{1}{|c|}{\ \ Time integration \ \ }&\multicolumn{1}{|c|}{\ \ Horowitz-Hubbeny \ \ }&\multicolumn{1}{|c|}{ \ \ Analytical \ \ } \\
      \hline
      $1$ & $1$ & $1$ & $-1.802098i$ & $-1.802177i$&$ -1.802177i  $ \\
      \hline
  $2$ & $1$ & $1$ & $-3.515213i$ & $-3.515213i$&$ -3.515213i$ \\
      \hline
  $3$ & $1$ & $1$ & $-5.228248i$ & $-5.228254i$&$ -5.228254i$ \\
      \hline
      $4$ & $1$ & $1$ & $-6.941137i$ & $-6.941296i$&$ -6.941296i$ \\
     \hline
     $5$ & $1$ & $1$ & $-8.654337i$ & $-8.654337i$&$ -8.654337i$ \\
     \hline
$1$ & $0$ & $1$ & $-1.713041i$ & $-1.713041i$&$ -1.713041i$ \\
      \hline
$1$ & $2$ & $1$ & $-2.072675i$ & $-2.069562i$&$ -2.069562i$ \\
      \hline
$1$ & $3$ & $1$ & $-2.551524i$ & $-2.515213i$&$ -2.515213i$ \\
     \hline
       $1$ & $4$ & $1$ & $-3.139086i$ & $-3.139124i$&$-3.139124i$ \\
      \hline
 $1$ & $5$ & $1$ & $-3.941332i$ & $-3.941296i$&$-3.941296i$ \\
      \hline
     $1$ & $1$ & $0$ & $-1.727273i$ & $-1.727273i$&$-1.727273i$ \\
      \hline
      $1$ & $1$ & $2$ & $-2.012838i$ & $-2.012756i$&$ -2.012756i$ \\
      \hline
 $1$ & $1$ & $3$ & $-2.012838i$ & $-2.326399i$&$ -2.326399i$ \\
     \hline
  $1$ & $1$ & $4$ & $-2.709162i$ & $-2.709208i$&$ -2.709208i$ \\
     \hline
     $1$ & $1$ & $5$ & $-3.135388i$ & $-3.135421i$&$ -3.135421i$ \\
          \hline
       \hline
  \end{tabular}\label{nueve}}
   \caption{\it Scalar quasinormal frequencies in $9$-dimensional $L=1$ Lifshitz black holes for various values of
$M$, $m$ and $k$. Results of calculation using time domain integration, Horowitz-Hubbeny method and exact analytical formula are included.}
      \end{table}
      \end{center}
\begin{center}
\begin{table}[htb!]
   \setlength{\belowcaptionskip}{5pt}  
  {\begin{tabular}{|c|c|c|c|c|c|}
      \hline
      \hline
       \multicolumn{1}{|c|}{\ \ $M$ \ \ }& \multicolumn{1}{|c|}{\ \ $k$ \ \ }& \multicolumn{1}{|c|}{\ \ $m$ \ \ }&
\multicolumn{1}{|c|}{\ \ Time integration \ \ }&\multicolumn{1}{|c|}{\ \ Horowitz-Hubbeny \ \ }&\multicolumn{1}{|c|}{ \ \ Analytical \ \ } \\
      \hline
      $1$ & $1$ & $1$ & $-1.819814i$ & $-1.819804i$&$ \ \ -1.819804i \ \ $ \\
      \hline
  $2$ & $1$ & $1$ & $-3.557720i$ & $-3.557627i$&$ -3.557627i$ \\
      \hline
  $3$ & $1$ & $1$ & $-5.295766i$ & $-5.295451i$&$-5.295451i$ \\
      \hline
      $4$ & $1$ & $1$ & $-7.034023i$ & $-7.033274i$&$ -7.033274i$ \\
     \hline
     $5$ & $1$ & $1$ & $-8.772563i$ & $-8.771098i$&$ -8.771098i$ \\
     \hline
$1$ & $0$ & $1$ & $-1.737823i$ & $-1.737823i$&$ -1.737823i$ \\
      \hline
$1$ & $2$ & $1$ & $-2.065770i$ & $-2.065745i$&$-2.065745i$ \\
      \hline
$1$ & $3$ & $1$ & $-2.475657i$ & $-2.475647i$&$ -2.475647i$ \\
     \hline
       $1$ & $4$ & $1$ & $-3.049581i$ & $-3.049510i$&$-3.049510i$ \\
      \hline
 $1$ & $5$ & $1$ & $-3.788032i$ & $-3.787333i$&$-3.787333i$ \\
      \hline
     $1$ & $1$ & $0$ & $-1.750000i$ & $-1.750000i$&$-1.750000i$ \\
      \hline
      $1$ & $1$ & $2$ & $-2.018032i$ & $-2.017985i$&$ -2.017985i$ \\
      \hline
 $1$ & $1$ & $3$ & $-2.317541i$ & $-2.317532i$&$-2.317532i$ \\
     \hline
  $1$ & $1$ & $4$ & $-2.688471i$ & $-2.688476i$&$ -2.688476i$ \\
     \hline
     $1$ & $1$ & $5$ & $-3.106290i$ & $-3.106289i$&$-3.106289i$ \\
          \hline
       \hline
  \end{tabular}\label{diez}}
   \caption{\it Scalar quasinormal frequencies in $10$-dimensional $L=1$ Lifshitz black holes for various values of
$M$, $m$ and $k$. Results of calculation using time domain integration, Horowitz-Hubbeny method and exact analytical formula are included.}
      \end{table}
      \end{center}
\begin{figure}[t]
\begin{center}
\scalebox{0.50}{\includegraphics{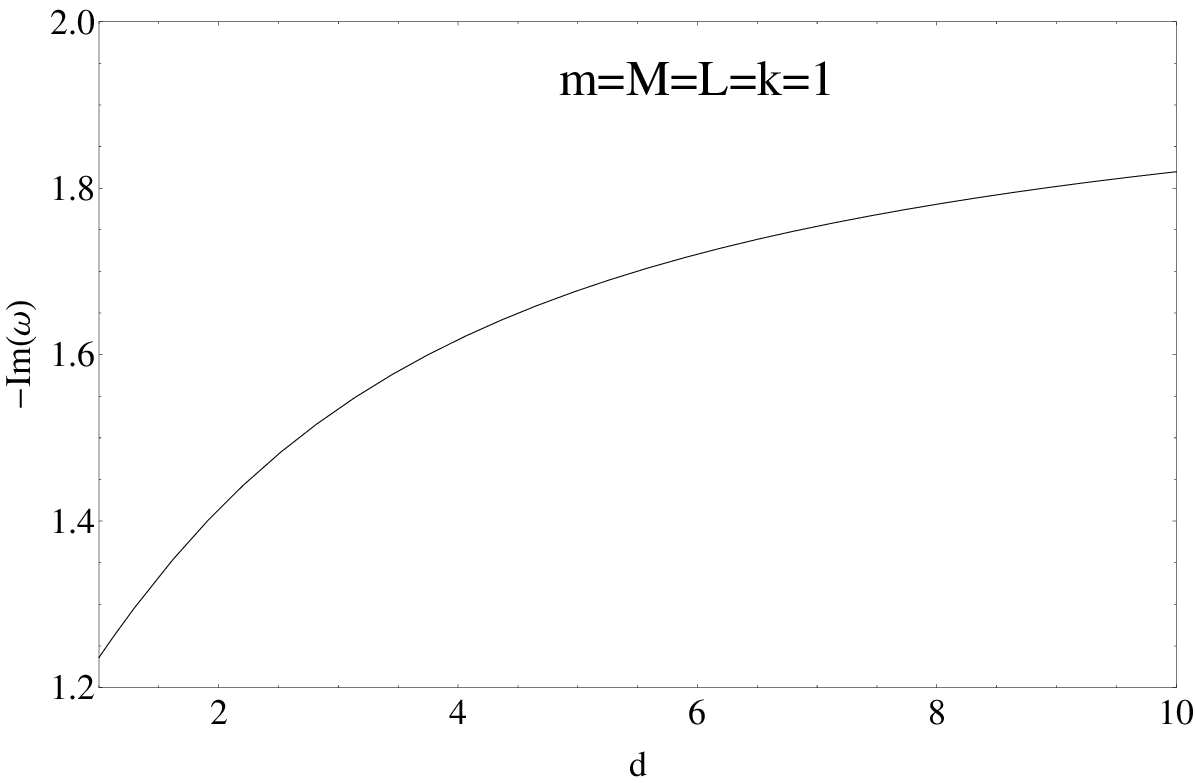}}
\end{center}
\caption{\it Dependence of scalar quasinormal frequencies with the dimensionality of Lifshitz black holes. The parameters of the solution
are $M=m=L=k=1$.} \label{fdepd}
\end{figure}
\begin{figure}[t]
\begin{center}
\scalebox{0.50}{\includegraphics{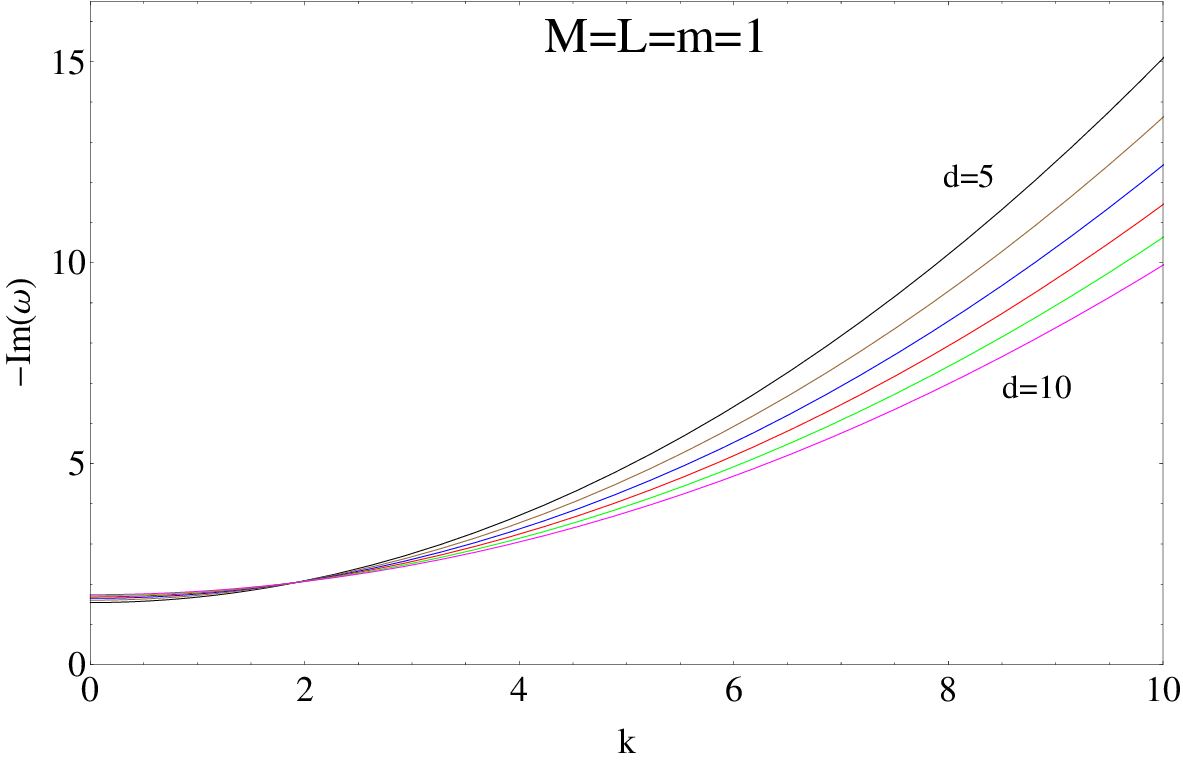}}
\end{center}
\caption{\it Dependence of scalar quasinormal frequencies with the wavevector $k$ for Lifshitz black holes of various dimensions. The
parameters of the solution are $M=m=L=1$.} \label{fdepk}
\end{figure}
\begin{figure}[t]
\begin{center}
\scalebox{0.50}{\includegraphics{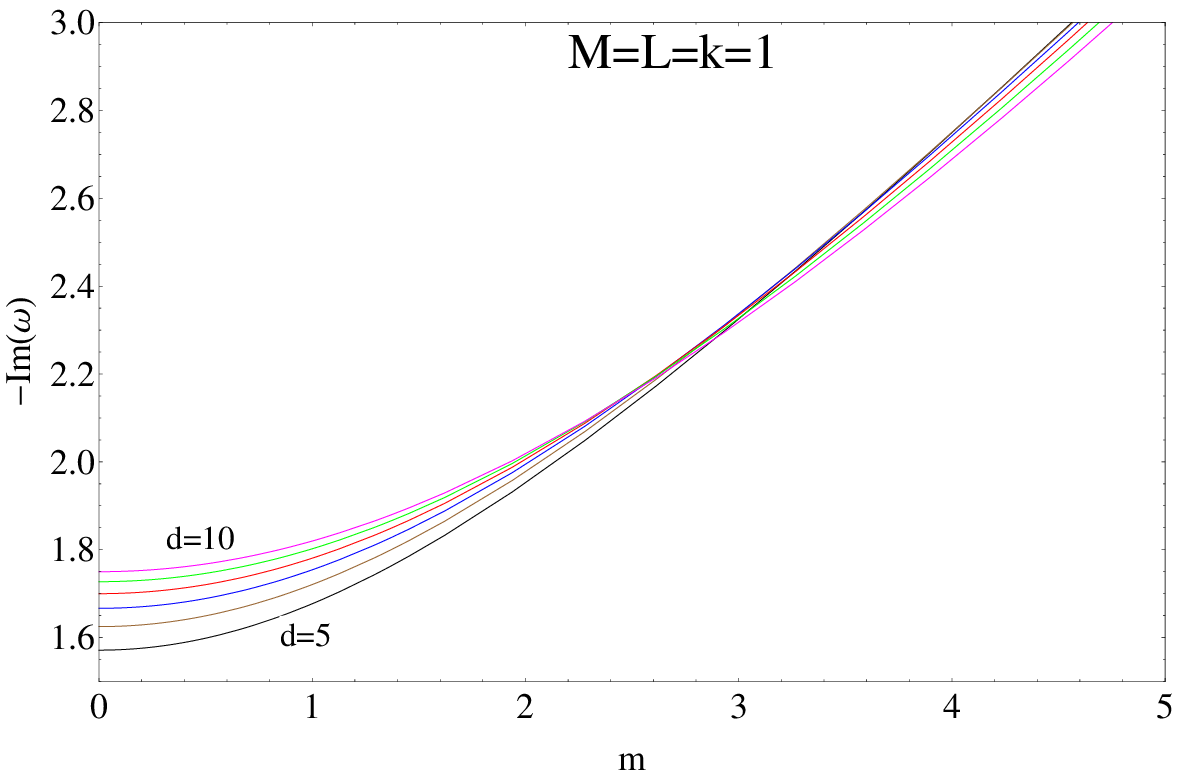}}
\end{center}
\caption{\it Dependence of scalar quasinormal frequencies with the field mass $m$ for Lifshitz black holes of various dimensions. The
parameters of the solution are $M=k=L=1$.} \label{fdepm}
\end{figure}
\begin{figure}[t]
\begin{center}
\scalebox{0.50}{\includegraphics{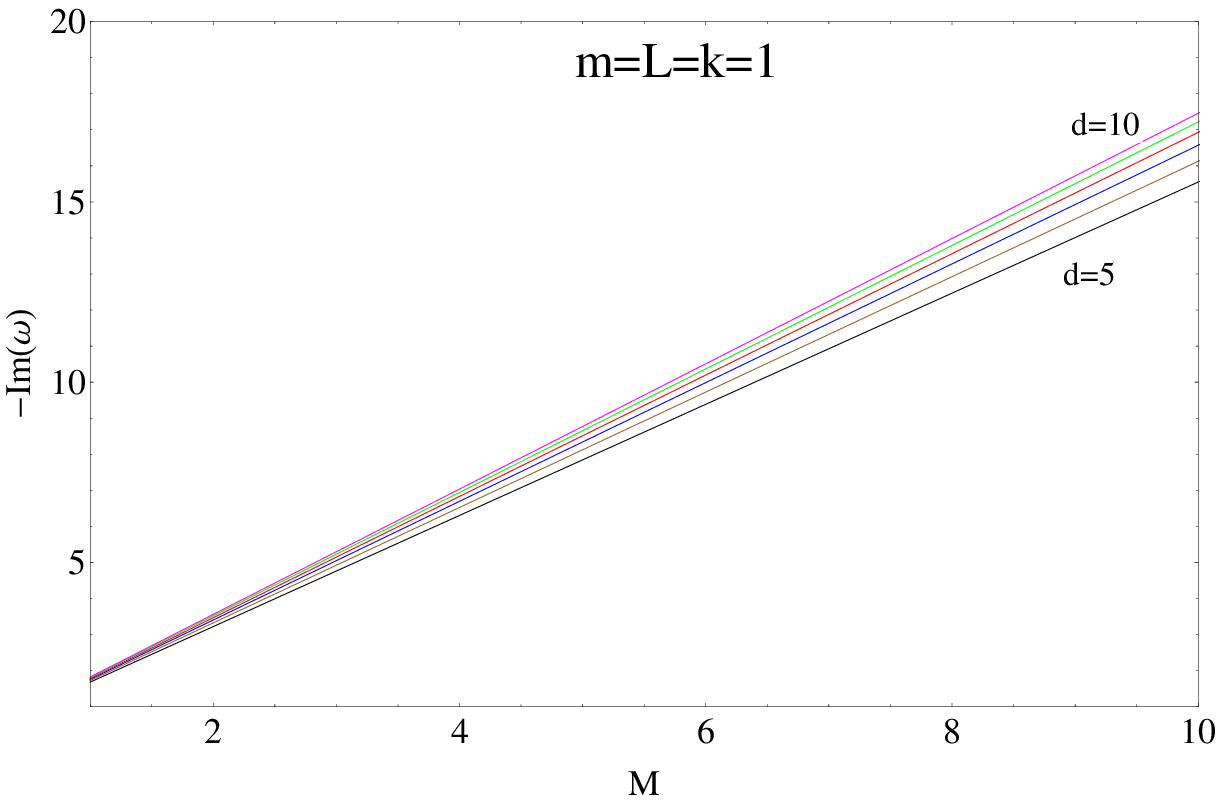}}
\end{center}
\caption{\it Dependence of scalar quasinormal frequencies with the black hole mass $M$ for Lifshitz black holes of various dimensions.
The parameters of the solution are $k=m=L=1$.} \label{fdepMM}
\end{figure}
\section{Exact quasinormal frequencies}
Equation (\ref{kgspherical3}) can be analytically solved, offering us the possibility of find exactly
the quasinormal frequencies of Lifzhitz black holes considered in this report.

In order to compute the quasinormal frequencies that dominated at intermediate times, we assume for the function $\varphi(t,r)$ in
equation (\ref{kgspherical3}) the time dependence
\begin{eqnarray}
\ \ \ \ \ \ \ \  \varphi(t,r)=Z(r)\exp(-i\omega t)
\end{eqnarray}
Then, for $z=2$ higher dimensional Lifshitz black holes whose metric tensor have the components given by (\ref{lbh2}),
the scalar field amplitude $Z(r)$  satisfies the equation
\begin{equation}\label{exact1}
{r}^{2} \left( {r}^{2}-M{L}^{2} \right)Z'' +r \left[\left( D+1 \right) {r}^{2}- \left( D
-1 \right) M{L}^{2} \right]Z'-\left( {L}^{4}{k}^{2}+{L}^{2}{r}^{2}{m}^{2}-{\frac {{\omega}^{2}{L}^{
6}}{{r}^{2}-M{L}^{2}}} \right) Z =0\quad ,
\end{equation}
were we denote the radial derivatives of a function $f(r)$ as $f'$ and drop from $\varphi$ in (\ref{kgspherical3}) the labels
$\mathbf{k}$ and $(E)$.

Introducing in the above equation the new variable $\zeta(r)$ defined as
\begin{equation}
\zeta=1-\frac{ML^{2}}{r^{2}}\quad ,
\end{equation}
such that $\zeta(r_{+})=0$ and $\zeta(\infty)=1$, we obtain
\begin{equation}\label{exact2}
    \zeta(\zeta-1)Z''(\zeta)+\frac{1}{2}\left[(D-4)\zeta+2 \right]Z'(\zeta)+{L}^{2}\left[ M{k}^{2}+\frac{{\omega}^{2}(\zeta-1)}{\zeta}
-\frac{{m}^{2}{M}^{2}}{\zeta-1}\right]Z(\zeta) =0
\end{equation}
Proposing a solution in the form $Z(\zeta)=\zeta^\alpha(1-\zeta)^\beta F(\zeta)$, where
\begin{eqnarray}\label{coef1}
\alpha&=&-\frac{iL\omega}{2M}  \,\\
\beta&=&\frac{1}{4}D+\frac{1}{4}\sqrt {4\,{m}^{2}{L}^{2}+{D}^{2}} \,\label{coef2}
\end{eqnarray}
we can transform (\ref{exact2}) into an hypergeometric equation
\begin{equation}\label{hyper1}\
\zeta(1-\zeta)F''(\zeta)+\left[c-(1+a+b)\zeta\right]F'(\zeta)-ab F(\zeta)=0~,
\end{equation}
whose general solution can be written as
\begin{equation}\label{hipersol1}
Z\left(\zeta\right) =\zeta^{\alpha}\left( 1-\zeta\right) ^{\beta}\left[ \lambda_{1}\
{}_{2}F_{1}\left( a,b,c,\zeta\right) +\lambda_{2}\ \zeta^{1-c}\ {}_{2}F_{1}\left(
b-c+1,a-c+1,2-c,\zeta\right) \right] \ ,
\end{equation}
where ${}_{2}F_{1}\left(
\zeta\right)$ is the hypergeometric function, and $\lambda_{1}$ and $\lambda_{2}$ are integration constants. The arguments of the
hypergeometric function are
\begin{align}\label{coef3}
a & =\alpha+\beta-\frac{3}{4}+\frac{1}{4M}\sqrt{\left( D-2\right) ^{2}M^{2}-4L^{2}\left(
Mk^{2}+\omega^{2}\right)} \quad , \\
b & =\alpha+\beta-\frac{3}{4}-\frac{1}{4M}\sqrt{\left( D-2\right) ^{2}M^{2}-4L^{2}\left(
Mk^{2}+\omega^{2}\right)} \quad , \\
c & =1+2\alpha\quad  .\label{coefc}
\end{align}
Taking into account (\ref{coefc}) we can write (\ref{hipersol1}) as
\begin{equation}\label{hipersol2}
Z\left(\zeta\right) =\lambda_{1}\zeta^{\alpha}\left( 1-\zeta\right) ^{\beta} \
{}_{2}F_{1}\left( a,b,c,\zeta\right) +\lambda_{2}\ \zeta^{-\alpha}\left( 1-\zeta\right) ^{\beta}\ {}_{2}F_{1}\left(
b-c+1,a-c+1,2-c,\zeta\right) \quad ,
\end{equation}
Ingoing boundary condition at the horizon $\zeta=0$ of the Lifshitz black holes fixes $\lambda_{2}=0$, then the general solution is
\begin{equation}\label{hipersol3}
Z\left( \zeta\right) =\lambda_{1}\zeta^{\alpha}\left( 1-\zeta\right) ^{\beta} \
{}_{2}F_{1}\left( a,b,c,\zeta\right)\quad .
\end{equation}
To take into account the boundary condition at spatial infinity $\zeta=1$, we can use the Kummer´s formula to connect
hypergeometric functions at $\zeta=0$ with that at $\zeta=1$,
\begin{eqnarray}\label{kummer}
\nonumber
{}_{2}F_{1}(a,b,c;\zeta)&=&\frac{\Gamma(c)\Gamma(c-a-b)}{\Gamma(c-a)\Gamma(c-b)}
{}_{2}F_1(a,b,a+b-c,1-\zeta)\\
&+&(1-\zeta)^{c-a-b}\frac{\Gamma(c)\Gamma(a+b-c)}{\Gamma(a)\Gamma(b)}F_1(c-a,c-b,c-a-b+1,1-z)\quad .
\end{eqnarray}
Using (\ref{coef2})-(\ref{coefc}) the above expression can be written as
\begin{eqnarray}\label{exact3}\
\nonumber
Z(\zeta) &=& \lambda_1 e^{\alpha \ln \zeta}(1-\zeta)^\beta\frac{\Gamma(c)\Gamma(c-a-b)}{\Gamma(c-a)\Gamma(c-b)} {}_{2}F_1(a,b,a+b-c,1-\zeta)\\
&+&\lambda_1 e^{\alpha \ln
\zeta}(1-\zeta)^{\frac{D}{2}-\beta}\frac{\Gamma(c)\Gamma(a+b-c)}{\Gamma(a)\Gamma(b)}{}_{2}F_1(c-a,c-b,c-a-b+1,1-\zeta)\quad .
\end{eqnarray}
Now, at $\zeta=1$ we have,
\begin{equation}\label{exactfin}\
Z(z\longrightarrow1) = \lambda_1 (1-\zeta)^\beta\frac{\Gamma(c)\Gamma(c-a-b)}{\Gamma(c-a)\Gamma(c-b)}+\lambda_1
(1-\zeta)^{\frac{D}{2}-\beta}\frac{\Gamma(c)\Gamma(a+b-c)}{\Gamma(a)\Gamma(b)}.
\end{equation}
The first term in the above expression tends to zero at spatial infinity, thus is consistent with the Dirichlet boundary condition
in this limit. In the following, we exclude terms that are present in hypergeometric functions considering only non integer values
of $c-a-b$. The power of $1-\zeta$ in the second term diverges for real scalar field masses and we need to impose that the quotient
of the Gamma functions in this term vanishes, allowing Dirichlet boundary conditions to be satisfied. This is true only if
either $\Gamma(a)$ or $\Gamma(b)$ diverges, what leaves us with two conditions that can be used in order to determine the
quasinormal frequencies, that is, either $a=-n$ or $b=-n$, where $n=0,1,2,...$.
Using (\ref{coef1}), (\ref{coef2}) and (\ref{coefc}) we find for the quasinormal frequencies the exact result
\begin{equation}\label{qnmf1}
    \omega_{n}=-\frac{i}{L}\frac{\left\{L^{2}k^{2}+M\left[(2n+1)\sqrt{D^{2}+4m^{2}L^{2}}+4n(n+1)+L^{2}m^{2}+5\right]\right\}}
    {\left[2(2n+1)+\sqrt{D^{2}+4m^{2}L^{2}}\right]}\quad .
\end{equation}
From the condition $b=-n$ we obtain the same result for the quasinormal frequencies. The above expression indicates that for real
scalar field masses all quasinormal frequencies are pure imaginary and negative, for all dimensions. Thus, higher dimensional Lifshitz
Black Holes are stable under real mass scalar fluctuations.

For imaginary masses, we can distinguish two cases. Putting $m^{2}\equiv-|m^{2}|$, if $\left(\frac{D}{2}\right)-|m^{2}|\geq0$,
we have $\beta>0$ and $\frac{D}{2}-\beta>0$. Thus, the powers of $1-\zeta$ in the two terms of equation (\ref{exactfin}) vanish
at spatial infinity for all values of $a$, $b$ and $c$, as required by Dirichlet boundary conditions, leading to a continuous
spectrum of quasinormal frequencies. For imaginary masses $\left(\frac{D}{2}\right)-|m^{2}|<0$ and there are waves present at
the spatial infinity, that frustrate the fulfillment of the quasinormal boundary conditions, thus we exclude this case from our analysis.

Using (\ref{qnmf1}) it is easy to confirm the results of the last section, as is seen in Figures \ref{fdepd} to \ref{fdepMM}, in
which we show the dependence of scalar quasinormal frequencies with the parameters involved in the perturbation problem, $D$, $k$,
$m$ and $M$. Again, the propagation of scalar waves outside higher dimensional Lifshitz black holes is a very simple phenomenon.

\section{Concluding remarks}
We have studied the propagation of a minimally coupled scalar field outside higher dimensional asymptotically Lifshitz black holes,
obtained from a theory that includes the more general quadratic corrections to Einstein gravity. We have solved the equation of
motion analytically for all dimensions and performed its numerical study, obtaining the time domain evolution and the quasinormal frequencies.

The quasinormal stage at late times is characterized by the absence of proper oscillations and the field is purely damped. We have
negative purely imaginary frequencies, implying complete stability of the black holes under scalar fluctuations. This picture is
the same for all dimensions independently of the variation of parameters that describe the background and the perturbations, such as
the black hole mass $M$, the scalar field mass $m$ and the wave-vector of the propagating fluctuation. As all these parameters increase
their values, Lifshitz black holes becomes more rigid, with a quicker decay of the probe field outside the event horizon.

The analytical result for the quasinormal spectrum shows us that the fundamental mode for any dimension is purely imaginary. Concerning the
gauge/gravity duality, such a result means that the holographic field theory has a small thermalization timescale. In the $z=3$ NMG case,
the same result has been found \cite{abdallaoliveira} and we conjecture that the purely imaginary quasinormal frequencies  are a universal
propriety of the black holes with Lifshitz scaling symmetry.

\section*{Acknowledgments}
This work has been supported by FAPESP (\emph{Funda\c c\~ao de,
Amparo \`a Pesquisa do Estado de S\~ao Paulo}), Brazil and the University of Cienfuegos, Cuba.
One of the authors (OPFP) express his gratitude to Professor Elcio Abdalla from the Department of Mathematical Physics at
University of S\~{a}o Paulo for the support during a research stay at his group, where this work was completed.


\end{document}